\documentclass[aps,pra,showpacs,amsmath,amssymb,amsfonts,lengthcheck,longbibliography,superscriptaddress,showpacs]{revtex4-1}

\usepackage{mathrsfs}  
\usepackage[usenames,dvipsnames]{color}
\usepackage{soul}
\usepackage{cancel}
\usepackage[
verbose,
colorlinks=true,
naturalnames=true,
linkcolor=blue,
]{hyperref}
\usepackage{graphicx}
\usepackage{subfigure}
\usepackage{amsthm}
\usepackage{verbatim}
\usepackage{dcolumn}% Align table columns on decimal point
\usepackage{bm}% bold math
\usepackage{epsf}

\hypersetup{colorlinks,linkcolor={blue},citecolor={blue},urlcolor={blue}} 

\newcommand{\bra}[1]{\left\langle #1\right|}
\newcommand{\ket}[1]{\left|#1\right\rangle}
\newcommand{\braket}[2]{\left\langle #1|#2\right\rangle}

\newcommand{\tr}[1]{\mathrm{tr}\left\{#1\right\}}

\newcommand{\la}{\left\langle}
\newcommand{\ra}{\right\rangle}

\newcommand{\e}[1]{\exp{\left(#1\right)}}
\newcommand{\lo}[1]{\ln{\left(#1\right)}}

\newcommand{\com}[2]{\left[#1,\,#2\right]}

\newcommand{\bla}{bla\\bla\\bla\\bla\\bla}

\newcommand{\mc}[1]{\mathcal{#1}}

\newcommand{\mf}[1]{\mathfrak{#1}}
\newcommand{\mrm}[1]{\mathrm{#1}}

\DeclareMathOperator*{\sumint}{%
	\mathchoice%
	{\ooalign{$\displaystyle\sum$\cr\hidewidth$\displaystyle\int$\hidewidth\cr}}
	{\ooalign{\raisebox{.14\height}{\scalebox{.7}{$\textstyle\sum$}}\cr\hidewidth$\textstyle\int$\hidewidth\cr}}
	{\ooalign{\raisebox{.2\height}{\scalebox{.6}{$\scriptstyle\sum$}}\cr$\scriptstyle\int$\cr}}
	{\ooalign{\raisebox{.2\height}{\scalebox{.6}{$\scriptstyle\sum$}}\cr$\scriptstyle\int$\cr}}
}

\newcommand{\eq}{\text{eq}}

\newcommand{\irr}{\text{irr}}

\renewcommand{\vec}[1]{\boldsymbol{#1}}

\begin{document}
	\title{Quantum and classical ergotropy from relative entropies}

	\author{Akira Sone}
	%\email{akira@aliroquantum.com}
	\affiliation{Aliro Technologies, Inc. Boston, Massachusetts 02135, USA}
	\affiliation{Theoretical Division, Los Alamos National Laboratory, Los Alamos, New Mexico 87545,USA}
	\affiliation{Center for Nonlinear Studies, Los Alamos National Laboratory, Los Alamos, New Mexico 87545, USA}
	
	\author{Sebastian Deffner}
	%\email{deffner@umbc.edu}
	\affiliation{Department of Physics, University of Maryland, Baltimore County, Baltimore, Maryland 21250, USA}
	\affiliation{Instituto de F\'isica `Gleb Wataghin', Universidade Estadual de Campinas, 13083-859, Campinas, S\~{a}o Paulo, Brazil}
	
	\begin{abstract}
		The quantum ergotropy quantifies the maximal amount of work that can be extracted from a quantum state without changing its entropy.  {Given that the ergotropy can be expressed as the difference of quantum and classical relative entropies of the quantum state with respect to the thermal state,  we define the classical ergotropy, which quantifies how much work can be extracted from distributions that are inhomogeneous on the energy surfaces.} A unified approach to treat both quantum as well as classical scenarios is provided by geometric quantum mechanics, for which we define the geometric relative entropy. The analysis is concluded with an application of the conceptual insight to conditional thermal states, and the correspondingly tightened maximum work theorem.
	\end{abstract}
	\maketitle
	\onecolumngrid
	
	\section{Introduction}
	According to its definition, the adjective \emph{ergotropic}  refers to the physiological mechanisms of a nervous system to favor an organism's capacity to expend energy \cite{Gellhorn1970}.  Generalizing this notion to physical systems, \emph{quantum ergotropy} was then coined to denote the maximal amount of work that can be extracted by isentropic transformations \cite{Allahverdyan04}. In particular, the quantum ergotropy quantifies the amount of energy that is stored in \emph{active} quantum states, and which can be extracted by making the state \emph{passive}~\cite{ Pusz1978,Koukoulekidis19, Gorecki80, Daniels81}.  In simple terms, a passive state is diagonal in the energy basis, and its eigenstates are ordered in descending magnitude of its eigenvalues. Gibbs states are then  called \emph{completely passive} \cite{Pusz1978}.
	
	The quantum ergotropy plays a prominent role in quantum thermodynamics \cite{DeffnerBook19}. In particular, when assessing the \emph{thermodynamic value} of genuine quantum properties \cite{Goold16,Marti17,Levy19,Santos19}, such as squeezed and nonequilibrium reservoirs \cite{Niedenzu2018,Cherubim2019}, coherence \cite{Francica20,Cakmak2020}, or quantum correlations \cite{Francica2017,Touil2021},  it has proven powerful.  However, if the quantum system is not in contact with a heat reservoir,  computing the quantum ergotropy is far from trivial. This is due to the fact that the ergotropy is determined by a maximum over all unitaries that can act upon the system \cite{Allahverdyan04}. Note that not all passive states can be reached by unitary operations, in particular, including the completely passive state. 
	
	{In this paper, given that the quantum ergotropy can be written as the difference of quantum and classical relative entropies (the Kullback--Leibler divergence of the eigenvalue distributions), we define a \emph{classical ergotropy}, which quantifies the maximal amount of work that can be extracted from inhomogeneities on the energy surfaces, which  have been shown to be analogous to quantum coherences \cite{Smith19Thesis,Smith2021}.}
	
	In a second part of the analysis, we turn to a unified framework, namely geometric quantum mechanics. Exploiting this approach \cite{Anza20a,Anza20b,Anza20c}, we  define the \emph{geometric relative entropy}.  With this, it becomes particularly transparent to characterize the one-time measurement approach to quantum work \cite{Deffner16,Beyer2020,Sone20a,Sone21b,Allahverdyan05}. In this paradigm,  work is determined by first measuring the energy of the system, and then letting it evolve under time-dependent dynamics. In contrast to the two-time measurement approach \cite{Kurchan01, Tasaki00,Talkner07,Huber2008,Campisi11,Deffner2011PRL,Kafri2012,Mazzola2013,Dorner2013,Roncaglia2014,Batalhao2014,An2015,Deffner2015PRL,Deffner2015PRE,Talkner16,Gardas2016,Bartolotta2018,Gardas2018,Touil2021PRQ}, no projective measurement is taken at the end of the process. Hence, the work probability distribution is entirely determined by the statistics \emph{conditioned} on the initial energy.  Here, we identify the distinct contributions to the thermodynamic cost of projective measurements by separating out the coherent and incoherent ergotropies, and the population mismatch in the conditional statistics. 
	
	Hence, by expressing the quantum ergotropy as a difference of relative entropies, we are able to (i) generalize the notion to classical scenarios,  and to (ii) elucidate the thermodynamics of projective measurements.  This analysis further cements ergotropy as one of the salient pillars of quantum thermodynamics. 
	
	{The paper is organized as follow. In Section~\ref{sec:QuantumErgotropy}, we review quantum ergotropy in terms of relative entropies and its relation to the quantum coherence in Section~\ref{sec:ErgotropyCoherence}. Then, we introduce the formulation of classical ergotropy in Section~\ref{sec:ClassicalErgotropy}, and discuss the geometric quantum mechanics approach to ergotropies in Section~\ref{sec:ErgotropyGeometry}. Finally, we discuss the physical meaning of the conditional thermal states in the second law of thermodynamics based on its ergotropy in Section~\ref{sec:ErgotropyConditionalThermal} before our conclusions in Section~\ref{sec:conclusion}. 
	}

	\section{Quantum Ergotropy}
	\label{sec:QuantumErgotropy}
	
	We begin by deriving a simple expression for the quantum ergotropy, which does not explicitly depend on the optimization over unitary maps.  To this end,  consider a quantum system with Hamiltonian $H$ and quantum state $\rho$. Then, the ergotropy is defined as ~\cite{Allahverdyan04}
	\begin{equation}
		\label{eq:ergo}
		\mc{E}\left(\rho\right) \equiv \tr{\rho\, H} - \min_{U\in\mc{U}}\left[ \tr{U\rho U^{\dagger}\,H}\right]\,,
	\end{equation}
	where $\mc{U}$ is the unitary group. 
	
	Our goal is now to express   Eq.~\eqref{eq:ergo} as a difference of relative entropies.  To this end, we write the quantum state $\rho$ in its ``ordered'' eigenbasis
	\begin{equation}
		\rho=\sumint_i p_i \ket{p_i}\bra{p_i}\quad\text{with}\quad  p_i \geq p_{i+1}\,.
	\end{equation}
	
	Let $\sigma$ be a second quantum state, which we write  
	\begin{equation}
		\sigma=\sumint_i s_i \ket{s_i}\bra{s_i}\quad\text{with}\quad  s_i \geq s_{i+1}\,.
	\end{equation}
	
	In principle, $\rho$ and $\sigma$ can be vastly different quantum states.  To better compare $\rho$ and $\sigma$,  it is then interesting to identify the unitary operation that takes $\rho$ as close as possible to $\sigma$.  Hence, considering the quantum relative entropy
	\begin{equation}
		S(U\rho U^\dagger||\sigma)\equiv\tr{\rho\lo{\rho}}-\tr{U\rho U^\dagger \lo{\sigma}}\,,
	\end{equation}
	it is known that the minimization of the quantum relative entropy over all the unitary operations is the classical relative entropy~\cite{Nielsen11} {(see Appendix~\ref{app:minQuantumRelativeEntropy} for the proof)} 
	\begin{equation}
		\min_{U\in\mc{U}}\left[S(U\rho U^{\dagger}||\sigma)\right]=\sumint_{i}p_i\lo{\frac{p_i}{s_i}}\equiv D\left(\rho||\sigma\right)\,.
		\label{eq:rel_c}
	\end{equation}
	
	To this end, we choose $\sigma$ as the Gibbs state
	\begin{equation}
		\rho^\mrm{eq}\equiv\frac{\e{-\beta H}}{Z} \quad\text{with}\quad  Z\equiv\tr{\e{-\beta H}}\,.
	\end{equation}
	
	For the sake of simplicity, we further assume that the eigenenergies are ordered in ascending magnitude, $E_i\leq E_{i+1}$. As an alternative expression, the quantum ergotropy can be expressed as the difference of relative entropies~\cite{Francica20,Lobejko21} {(see Appendix~\ref{app:ErgotropyRelative} for the proof)} 
	\begin{equation}
		\label{eq:ergo_2}
		\beta\, \mc{E}(\rho)=S(\rho||\rho^\mrm{eq})-D(\rho||\rho^\mrm{eq})\,.
	\end{equation}
	
	Note that the quantum ergotropy does not depend on the specific value of the temperature, but
	rather   Eq.~\eqref{eq:ergo_2} holds for any  $\beta$.
	In conclusion, the quantum ergotropy is written as the difference of the quantum and classical relative entropies of the quantum state $\rho$ with respect to $\rho^\mrm{eq}$.  Note that   Eq.~\eqref{eq:ergo_2} is entirely determined by $\rho$ and  $\rho^\mrm{eq}$, and independent of any optimization.
	
	\section{Ergotropy from Quantum Coherence}
	\label{sec:ErgotropyCoherence}
	
	It was recently recognized \cite{Francica20,Cakmak2020,Touil2021} that the quantum ergotropy \eqref{eq:ergo} can be separated into two fundamentally different contributions  
	\begin{equation}
		\mc{E}(\rho)=\mc{E}_i(\rho)+\mc{E}_c(\rho)\,.
	\end{equation}
	
	The \emph{incoherent ergotropy} $\mc{E}_i(\rho)$ denotes the maximal work that can be extracted from $\rho$ without changing its coherence, which is defined as ~\cite{Francica20} 
	\begin{equation}
		\mc{E}_i(\rho)\equiv \tr{(\rho-\tau)H}\,.
	\end{equation}
	
	Here, we call $\tau$ the coherence-invariant state of $\rho$, which is defined as~\cite{Francica20} 
	\begin{equation}
		\tr{\tau H}=\min_{U\in\mc{U}^{(i)}}\tr{U\rho U^{\dagger}H}\,,
	\end{equation}
	where $\mc{U}^{(i)}$ is the set of unitary operations without changing the coherence of $\rho$. Refer to Ref.~\cite{Francica20} for more details about $\mathcal{U}^{(i)}$.
	
	The \emph{coherent ergotropy} $\mc{E}_c(\rho)$ is the work that is exclusively stored in the coherences. This can be quantified by the relative entropy of coherence~\cite{Baumgratz14}  
	\begin{equation}
		\mc{C}(\rho)=\mc{H}\left(\mc{L}(\rho)\right)-\mc{H}(\rho)\,,
	\end{equation}
	where {$\mc{H}(\rho)\equiv-\tr{\rho\lo{\rho}}$ is the von Neumann entropy of $\rho$}, and $\mc{L}$ is the purely dephasing map, i.e., the map that removes all coherences but leaves the diagonal elements in the energy basis invariant.   {From the expression of the coherent ergotropy derived in Ref.~\cite{Francica20} and   Eq.~\eqref{eq:rel_c}}, the coherent ergotropy can be rewritten in terms of classical relative entropy
	\begin{equation}
		\label{eq:coh}
		\beta\,\mc{E}_c(\rho)=\mc{C}(\rho)+S(\mc{L}(\tau)||\rho^\mrm{eq})-D(\rho||\rho^\mrm{eq})\,.
	\end{equation}
	
	Hence, we conclude that there are three distinct contributions to the coherent ergotropy. Namely, work can be extracted not only from the coherences directly, but also from the population mismatch between the completely decohered state and the corresponding thermal state.  However, the total extractable work is lowered by the fact that generally $\rho$ is not diagonal in energy; hence, the classical relative entropy is different from the quantum relative entropy of the completely decohered state.
	
	\section{Classical Ergotropy from Inhomogeneity}
	\label{sec:ClassicalErgotropy}
	
	Remarkably, the above discussion of the quantum treatment can be generalized to purely classical scenarios. It was recently recognized that distributions that are inhomogeneous on the energy surfaces can be considered the classical equivalent of quantum states with coherences \cite{Smith19Thesis,Smith2021}.  Therefore, we proceed by defining the \emph{classical ergotropy}, which quantifies the maximal work that can be extracted from inhomogeneous distributions under Hamiltonian dynamics, i.e., under the classical equivalent of unitary maps.
	
	We start with the classical distribution, $p(\Gamma)$, which measures how likely it is to find a system at a point in phase space $\Gamma$. Now consider a situation in which $\Gamma$ is sampled microcanonically from an (initial) energy surface $A$; we then let $p_A(\Gamma)$ evolve under Liouville's equation.  We are interested in assessing how close to equilibrium the system is driven. To this end,  consider the joint   {distribution} of finding $\Gamma'$ on energy surface $B$, given that $\Gamma$ was sampled from energy surface $A$  
	\begin{equation}
		p_{B|A}(\Gamma,\Gamma' ) = p(\Gamma' |\Gamma)\,p_A(\Gamma)\,,
	\end{equation}
	where $p(\Gamma' |\Gamma)$ is the classical transition probability distribution, which satisfies
	\begin{equation}
		\label{eq:norm}
		\int d\Gamma\, p(\Gamma' |\Gamma)=\int d\Gamma'\, p(\Gamma' |\Gamma)=1\,,
	\end{equation}
	which follows from Liouville's theorem and normalization. In the following, we formulate the classical ergotropy by focusing on the joint distribution $p_{B|A}(\Gamma,\Gamma')$ in general classical systems.

	In complete analogy to the quantum case, we now consider the relative entropy of $p_{B|A}(\Gamma,\Gamma' )$ with respect to the thermal distribution on energy surface $B$   
	\begin{equation}
		p_B^\mrm{eq}(\Gamma')=\frac{\e{-\beta E_B(\Gamma')}}{Z} \quad\text{with}\quad Z\equiv\int d\Gamma' \e{-\beta E_B(\Gamma')}\,.
	\end{equation}
	
	We can write
	\begin{equation}
		\label{eq:relative}
		D(p_{B|A}||p_B^\mrm{eq})=\int d\Gamma\int d\Gamma'\, p_{B|A}(\Gamma,\Gamma') \lo{\frac{p_{B|A}(\Gamma,\Gamma')}{p^\mrm{eq}_B(\Gamma')}}\,.
	\end{equation}
	
	Eq.~\eqref{eq:relative} is a divergence-like quantity, which becomes non-negative only for the thermodynamic scenario {(See Appendix~\ref{app:divergence-like})}.
	Note that the normalization of the transition probabilities \eqref{eq:norm} is essential to guarantee that the classical distributions, $p_{B|A}$ and $p_B^\mrm{eq}$, have the same support.  
	
	As before, we then seek a ``transformed'' joint distribution $\mathscr{Q}_{B|A}$ for which the relative entropy $D\left(\mathscr{Q}_{B|A}||p_B^{\eq}\right)$ becomes minimal.  This $\mathscr{Q}_{B|A}$  can be written as 
	\begin{equation}
		\mathscr{Q}_{B|A}(\Gamma'' ,\Gamma)\equiv \int d\Gamma'\, q(\Gamma'' |\Gamma' )p(\Gamma' |\Gamma)p_A(\Gamma)\,,
	\end{equation}
	and we need to minimize $D\left(\mathscr{Q}_{B|A}||p_B^{\eq}\right)$ as a function of the transition probability   {distribution}  $q(\Gamma'' |\Gamma' )$. We start by recognizing that the convolution of two transition   {probability distributions} is also a transition probability   {distribution}
	\begin{equation}
		\xi(\Gamma'' |\Gamma)\equiv\int d\Gamma'\,  q(\Gamma'' |\Gamma' )p(\Gamma' |\Gamma)\,.
	\end{equation}
	
	Then, we have $\mathscr{Q}_{B|A}(\Gamma'' ,\Gamma) = \xi(\Gamma'' |\Gamma)\,p_A(\Gamma)$, thus we need to minimize $D\left(\mathscr{Q}_{B|A}||p_B^{\eq}\right)$ as a function $\xi$. In a complete analogy to the quantum case, we can choose $\xi(\Gamma'|\Gamma)=\delta(\Gamma'-\Gamma)$ and obtain the following result~(see Appendix~\ref{app:MinClassicalRelativeEntropy} for the proof) 
	\begin{equation}
		\min_{\xi}\left[D\left(\mathscr{Q}_{B|A}||p_B^{\eq}\right)\right] = D\left(p_A||p_B^{\eq}\right)\,.
		\label{eq:MinClassicalRelativeEntropy}
	\end{equation}
	
	Accordingly,  we define the classical ergotropy  as 
	\begin{equation}
		\label{eq:ergo_3}
		\beta\,\mc{E}_\mrm{class}(p_{B|A})\equiv D\left(p_{B|A}||p_B^{\eq}\right)-D\left(p_A||p_B^{\eq}\right)\,,
	\end{equation}
	which quantifies the maximal amount of work that can be extracted from the joint distribution $p_{B|A}$ under Liouvillian maps. Remarkably,  both the quantum \eqref{eq:ergo_2} as well as the classical \eqref{eq:ergo_3} ergotropy comprise the classical relative entropy with respect to a thermal state. 
	
	Eq.~\eqref{eq:ergo_3} can also be re-written to resemble more closely the established expression of the quantum ergotropy \eqref{eq:ergo}.  We have
	\begin{equation}
		\mc{E}_\mrm{class}(p_{B|A}) = \int d\Gamma'\,\varphi_B(\Gamma')E_B(\Gamma')
	\end{equation}
	where we introduced
	\begin{equation}
		\varphi_B(\Gamma') = \int d\Gamma\, p_{B|A}(\Gamma',\Gamma)-p_A(\Gamma')\,.
	\end{equation}
	
	In this form, it becomes apparent that the classical ergotropy quantifies the maximal amount of work stored in the inhomogeneities. Notice that  $\varphi_B$ is not an explicit function of the Hamiltonian of the system, which was shown to be a classical equivalent of the quantum coherences \cite{Smith19Thesis,Smith2021}. This is the analogy of how the quantum ergotropy quantifies the maximal work extractable from quantum coherences.

	\section{Ergotropy in Geometric Quantum Mechanics}
	\label{sec:ErgotropyGeometry}
	
	Thus far, we have seen that in quantum as well as in classical systems,  work can be extracted by ``reshaping'' the states in phase space without changing their entropy. Remarkably, in either case, the ergotropy is given by a difference of relative entropies (see   Eqs.~\eqref{eq:ergo_2} and \eqref{eq:ergo_3}). The natural question arises as to whether the quantum-to-classical limit can be taken systematically, or rather the seemingly independent results can be derived within a unifying framework.
	
	Only very recently, Anza and Crutchfield \cite{Anza20a,Anza20b,Anza20c} recognized that for such thermodynamic considerations, so-called geometric quantum mechanics \cite{Ashtekar99,Zyczkowski17,Marmo07} are a uniquely suited paradigm. In standard quantum theory,  a quantum state  is described by a density operator $\rho$, which can be expanded in many different decompositions of pure states.   However, an often overlooked consequence is that, thus,  the probabilistic interpretation of quantum states is not unique.  To remedy this issue,  \emph{geometric quantum states} \cite{Ashtekar99,Zyczkowski17,Marmo07} were introduced, which are probability distributions on the manifold spanned by the quantum states. In this sense, classical and quantum mechanics only differ in the geometric properties of the underlying manifold.
	
	We proceed by briefly outlining the main notions of geometric quantum mechanics, which is well developed (cf. Refs.~ \cite{Ashtekar99,Zyczkowski17,Marmo07,Anza20a,Anza20b,Anza20c} for a more complete exposition). In the geometric approach, a pure quantum state $\ket{\psi}$ is described as a point in a complex projective space $\mc{V}_d\equiv\mathbb{C}P^{d-1}$~\cite{Zyczkowski17}, where $d$ is the dimension of the Hilbert space. Note that $d$ can also be infinite~\cite{Ashtekar99}. Here,  $\vec{z}$ is the set of complex homogeneous coordinates in $\mc{V}_d$, and $\vec{z^*}$ is the complex conjugate. 
	
	Hence, any pure state $\ket{\psi}$ can be written as 
	\begin{equation}
		\ket{\psi(\vec{z})}=\sum_{\alpha=0}^{d-1}z_{\alpha}\ket{e_{\alpha}}\,,  
	\end{equation}
	where $\{e_{\alpha}\}_{\alpha=0}^{d-1}$ is an arbitrary basis. The geometry of the manifold is determined by the Fubini--Study metric \cite{Zyczkowski17} 
	\begin{equation}
		ds^2= 2\sum_{\alpha,\gamma}g_{\alpha\gamma^*}dz_{\alpha}dz^*_{\gamma}\equiv  \frac{1}{2}\sum_{\alpha,\gamma}\partial_{z_{\alpha}}\partial_{z^*_{\gamma}}\lo{\vec{z}\cdot\vec{z^*}}dz_{\alpha}dz^*_{\gamma}\,,
	\end{equation}
	where we define $g_{\alpha\gamma^{*}}\equiv\frac{1}{4}\partial_{z_{\alpha}}\partial_{z_{\gamma}}\ln(\vec{z}\cdot\vec{z}^{*})$ and which allows to define a unique, unitarily invariant volume element,  $dV \equiv\sqrt{\det(g)}\,d\vec{z}d\vec{z^*} $.
	
	It is easy to recognize that pure states are represented as generalized delta functions on the projective space.  In particular, for $\ket{\psi_0}\equiv\ket{\psi(\vec{z}_0)}$, the corresponding geometric quantum state becomes
	\begin{equation}
		\mc{P}(\vec{z}) = \widetilde{\delta}(\vec{z}-\vec{z}_0)\equiv\frac{\delta(\vec{z}-\vec{z}_0)}{\sqrt{\det(g)}}\,,
	\end{equation}
	where we introduce the coordinate-covariant Dirac delta.  Any (mixed) quantum state can  then be written  as 
	\begin{equation}
		\rho=\int_{\mc{V}_d}dV\,  \mc{P}(\vec{z})\,\ket{\psi(\vec{z})}\bra{\psi(\vec{z})}\,,
	\end{equation}
	where the geometric quantum states are given by
	\begin{equation}
		\mc{P}(\vec{z})=\sum_{j=1}^{d}p_j\,\widetilde{\delta}\left(\vec{z}-\vec{z}_j^p\right)\,,
	\end{equation}
	and $p_j$ is again the eigenvalues of $\rho$ and  $\vec{z}_j^p\equiv\vec{z}(\ket{p_j})$.
	
	We are now equipped to return to the expressions for the quantum and classical ergotropies,   Eqs.~\eqref{eq:ergo_2} and \eqref{eq:ergo_3}, respectively. We immediately recognize that to proceed, we have to consider a generalization of the relative entropy to geometric quantum states. In complete analogy to the classical case, we need to guarantee that the geometric quantum states have the same support \cite{Anza21}.  Hence, we introduce a geometric quantum generalization of the conditional distribution to include a generalized transition probability   {distribution}.  To this end,  consider
	\begin{equation}
		\widetilde{\mc{P}}(\vec{z}) \equiv \sum_{j=1}^{d} p_j\,\widetilde{\delta}\left(\vec{z}-\vec{z}_j^s\right)\,,
		\label{eq:mathscrPdef}
	\end{equation}
	where now $\vec{z}_j^s\equiv\vec{z}(\ket{s_j})$, and $\ket{s_j}$ is an eigenstate of a density operator $\sigma$. The density operator, $\widetilde{\rho}$, corresponding to $\widetilde{\mc{P}}(\vec{z})$ reads
	\begin{equation}
		\widetilde{\rho}= \widetilde{U}\,\rho\, \widetilde{U}^{\dagger}  = \sum_j p_j\,\ket{s_j}\bra{s_j}\,,
	\end{equation}
	where $\widetilde{U}$ is the ``optimal'' unitary maps.
	
	The \emph{geometric relative entropy} is then defined as
	\begin{equation}
		\mc{D}\left(\widetilde{\mc{P}}||\mc{S}\right) \equiv \int_{\mc{V}_d} dV\, \widetilde{\mc{P}}(\vec{z})\lo{\frac{\widetilde{\mc{P}}(\vec{z})}{\mc{S}(\vec{z})}}\,,
	\end{equation}
	where $\mc{S}$ is the geometric quantum state corresponding to $\sigma$ (same as before). Moreover, we have by construction $\mc{D}\left(\widetilde{\mc{P}}||\mc{S}\right)= S\left(\widetilde{\rho}||\sigma\right)=D\left(\rho||\sigma\right)$,  and we conclude that the geometric relative entropy is identical in value to the classical relative entropy \eqref{eq:rel_c}.  Therefore, we can write the quantum ergotropy \eqref{eq:ergo_2} as
	\begin{equation}
		\label{eq:ergo_4}
		\beta \,\mc{E}(\rho)=S(\rho||\rho^\mrm{eq})-\mc{D}\left(\widetilde{\mc{P}}||\mc{P}^\mrm{eq}\right)\,,
	\end{equation}
	where $\mc{P}^\mrm{eq}$ is the geometric quantum state corresponding to $\rho^\mrm{eq}$. In other words, the quantum ergotropy is the difference of the relative entropies of the density operator and the geometric quantum state with respect to $\rho^\mrm{eq}$.
	
	Remarkably, also the classical case can be fully treated within the geometric approach. To this end, note that for any classical distribution, we can construct the corresponding geometric quantum state.  Therefore, it now becomes a fair comparison to consider the difference of quantum and classical ergotropy, $\Delta \mc{E}\equiv \mc{E}(\rho)-\mc{E}_\mrm{class}(\rho)$. It is not far-fetched to realize that $\Delta \mc{E}$ is the genuinely quantum contribution to the extractable work.  A more careful analysis of this contribution may be related to quantum correlations (see also Ref.~\cite{Touil2021}), yet a thorough analysis is beyond the scope of the present discussion. Rather, the remainder of this analysis is dedicated to an application of the gained insight to quantum work relations.

	\section{Ergotropy from Conditional Thermal States}
	\label{sec:ErgotropyConditionalThermal}
	
	To this end, imagine a closed system that is driven by the variation of some external control parameter.  We denote the initial Hamiltonian by $H_A$ and the final Hamiltonian by $H_B$, and the average work is simply given by $\la W\ra=\la H_B\ra-\la H_A\ra$.  The maximum work theorem predicts that $\la W\ra$ is always larger than the work performed for quasistastic \mbox{driving \cite{DeffnerBook19}}. If the system was initially prepared in a thermal state, the quasistatic work is nothing but the difference in Helmholtz free energy $\Delta F$ \cite{Deffner2010PRL}. The difference of total work and free energy difference is called \emph{irreversible work}, and we have \mbox{$\la W_\mrm{irr}\ra=\la W\ra-\Delta F\geq 0$ \cite{Deffner2010PRL}}.  Only rather recently, it was recognized that a sharper inequality can be derived, for both \mbox{quantum \cite{Deffner16}} as well as classical \cite{Sone21b} systems if the quantum work statistics are conditioned on the initial state.  Note that this corresponds to the one-time measurement approach, where only one projective measurement is taken at the beginning of the process. 
	
	In particular, the following was shown \cite{Deffner16,Sone21b} 
	\begin{equation}
		\beta\la W_{\irr}\ra \geq S\left(\varrho_B||\rho_B^{\eq}\right)\,, 
		\label{eq:2ndLaw}
	\end{equation}
	where  $\rho_B^{\eq}=\e{-\beta H_B}/Z_B$, and $\varrho_B$ is called the conditional thermal state \cite{Sone21b}. It reads~\cite{Deffner16}
	\begin{equation}
		\varrho_B\equiv \sum_j \frac{\e{-\beta\, h_B(j_A)}}{\mc{Z}(B|A)}\,U_{\tau}\ket{j_A}\bra{j_A}U_{\tau}^{\dagger}\,,
		\label{eq:guessedstate}
	\end{equation}
	where $\ket{j_A}$ is an eigenstate of the initial Hamiltonian $H_A$. Further, $U_\tau$ is the unitary evolution operator corresponding to driving the system from $H_A$ to $H_B$, and 
	\begin{equation}
		h_B(j_A)\equiv \bra{j_A}U_{\tau}^{\dagger}H_B U_{\tau}\ket{j_A}\,.
	\end{equation}
	
	{Finally,  $\mc{Z}(B|A)$ is the conditional partition function of $\varrho_B$. Since the discovery of \mbox{  Eq.~\eqref{eq:2ndLaw}}, the significance of the conditional thermal state has been somewhat obscure.   In Ref.~\cite{Deffner16,Sone20a}, the lower bound in   Eq.~\eqref{eq:2ndLaw} was understood as some contribution to the usable work that would have been destroyed by a second projective measurement. Yet, a transparent interpretation is lacking.}
	
	Remarkably, it is not hard to see that $\varrho_B$ is a representation of the \emph{geometric canonical ensemble} as proposed by Anza and Crutchfield \cite{Anza20a,Anza20c}.  The geometric canonical ensemble is defined as the geometric state that maximizes the corresponding Shannon entropy under the usual boundary conditions \cite{Jaynes57}. Specifically, we have~ \cite{Anza20a,Anza20c} \begin{equation}
		\mf{P}(\vec{z})\equiv \frac{\e{-\beta\, h(\vec{z})}}{\mc{Z}}\,,
		\label{eq:GcanonicalEnsemble}
	\end{equation}
	where $h(\vec{z})\equiv \bra{\psi(\vec{z})}H\ket{\psi(\vec{z})}$ and the geometric partition function is 
	\begin{equation}
		\mc{Z}\equiv \int_{\mc{V}_d}dV  \e{-\beta\, h(\vec{z})}\,.
	\end{equation}
	
	Thus, to maintain the consistency of the presentation, we continue to employ the geometric formulation of quantum states.
	Now, consider the geometric representation of $\varrho_B$ 
	\begin{equation}
		\varrho_B=\int_{\mc{V}_d} dV \mf{P}_B\,(\vec{z})\,\ket{\psi(\vec{z})}\bra{\psi(\vec{z})}\
	\end{equation}
	and we have
	\begin{equation}
		\mf{P}_B(\vec{z}) =\sum_j \frac{\e{-\beta\, h_B(\vec{z})}}{\mc{Z}(B|A)}\, \widetilde{\delta}\left(\vec{z}-\vec{z}_j\right)\,,   
		\label{eq:GstateConditionalThermal}
	\end{equation}
	where, as before, $h_B(\vec{z})\equiv\bra{\psi(\vec{z})}H_B\ket{\psi(\vec{z})}$ and the covariant Dirac delta is evaluated at $\ket{\psi(\vec{z}_j)}\equiv U_{\tau}\ket{j_A}$.  Comparing   Eqs. \eqref{eq:GcanonicalEnsemble} and \eqref{eq:GstateConditionalThermal}, we immediately recognize that the $\mf{P}_B(\vec{z})$ is nothing but the geometric canonical state evaluated on the quantum manifold. 
	
	The natural question arises as to whether any work can be extracted from the geometric ensemble. To this end, consider the corresponding ergotropy \eqref{eq:ergo_4} 
	\begin{equation}
		\beta\,\mc{E}(\varrho_B) = S\left(\varrho_B||\rho_B^{\eq}\right) - \mc{D}\left(\widetilde{\mf{P}}_B||\mf{P}_B^{\eq}\right)\,,
		\label{eq:ErgotropyGeometricCanonicalEnsemble}
	\end{equation}
	where, in complete analogy to the above, $\widetilde{\mf{P}}_B$ is given by
	\begin{equation}
		\widetilde{\mf{P}}_B(\vec{z})\equiv \sum_j\frac{\e{-\beta h_B(j_A)}}{\mc{Z}(B|A)}\,\widetilde{\delta}\left(\vec{z}-\vec{z}_j^{\eq}\right)\,, 
	\end{equation}
	and now $\vec{z}_j^{\eq}\equiv\vec{z}\left(\ket{j_B}\right)$, where $\ket{j_B}$ is the eigenstate of the final Hamiltonian $H_B$. Thus,  exploiting   Eq.~\eqref{eq:coh}, we can write the sharpened maximum work theorem \eqref{eq:2ndLaw} as 
	\begin{equation}
		\beta\la W_{\text{irr}}\ra \geq \beta\mc{E}_{i}(\varrho_B)+\mc{C}(\varrho_B)+ S\left(\mc{L}(\tau_B)||\rho_B^{\eq}\right)\,,
		\label{eq:2ndLawCoherence}
	\end{equation}
	where $\tau_B$ is the coherence-invariant state of $\varrho_B$.
	In conclusion, realizing that the conditional thermal state \eqref{eq:guessedstate} is nothing but a representation of the geometric canonical ensemble, the physical interpretation of the sharpened maximum work theorem \eqref{eq:2ndLaw} becomes apparent. The lower bound on the irreversible work has three contributions, namely the incoherent ergotropy and the quantum coherences stored in the conditional thermal state, and the population mismatch between $\varrho_B$ and $\rho_B^{\eq}$.  Therefore, we conclude that the conditional thermal state provides an informational contribution from its coherence to the second law.
	From the fact that the classical and quantum ergotropy share the same geometric relative entropy, we emphasize that thermodynamics based on geometric quantum mechanics is a unified approach to the quantum-to-classical limit.

	\section{Conclusions}
	\label{sec:conclusion}
	In conclusion, motivated by the desire to express the maximally extractable work in a form independent of the optimization over unitary operations, we have obtained several results. {Given that the quantum ergotropy can be expressed as the difference of the quantum and classical relative entropies, we identified three distinct contributions to the coherent ergotropy,  of which the relative entropy of coherence and the population mismatch between thermal state and fully decohered state are the most important.} This insight was extended to classical systems, in which inhomogeneities in the energy distribution play the role of quantum coherences.  To quantify how much work can be extracted from classical states, we introduced the classical ergotropy, and we postulated that the genuine quantum contribution to the ergotropy is given by the difference of the quantum and classical expressions. Our analysis provides a consistent approach to maximum work extraction in both quantum and
	classical systems. In particular, we have not only shown that classical inhomogeneities play the role of ``classical coherence", but also that work can be extracted that is quantified by the classical ergotropy. This was solidified by exploiting the geometric approach to quantum mechanics, in which quantum and classical states can be treated in a unified framework. As an application, we demonstrated that the recently introduced notion of ``conditional thermal state'' actually belongs to the family of geometric canonical ensembles and that, hence, the corresponding sharpened maximum work theorem becomes easy to interpret. This demonstrates that understanding quantum as well as classical ergotropies is an essential pillar of modern thermodynamics with a myriad of potential applications. Finally, these results demonstrate that the geometric approach can be regarded as a methodology of unifying the quantum and classical approaches to the second law of thermodynamics.

	\acknowledgements{We would like to thank Fabio Anza, Christopher Jarzynski, and Kanupriya Sinha for insightful discussions. A.S. was supported by the U.S. Department of Energy, the Laboratory Directed Research and Development (LDRD) program and the Center for Nonlinear Studies at LANL. He is now supported by the internal R\&D from Aliro Technologies, Inc. S.D. acknowledges support from the U.S. National Science Foundation under Grant No. DMR-2010127.}

	\appendix

	\section{Proof of   Eq.~\eqref{eq:rel_c}}
	\label{app:minQuantumRelativeEntropy}
	\begin{proof}
		In this section, we provide an alternative proof of   Eq.~\eqref{eq:rel_c} different from the method in Ref.~\cite{Nielsen11}. Actually, the procedure of the proof is similar to that of quantum ergotropy, which was introduced in Ref.~\cite{Allahverdyan04}.
		
		Let us consider the quantum relative entropy
		\begin{equation}
			S(U\rho U^\dagger||\sigma)\equiv\tr{\rho\lo{\rho}}-\tr{U\rho U^\dagger \lo{\sigma}}\,.
		\end{equation}
		
		In order to identify the specific $U$ such that $S(U\rho U^\dagger||\sigma)$ is minimized, we now parametrize a variation as $\delta U=(\delta X) U$, where $\delta X$ is an arbitrary infinitesimal, anti-Hermitian operator, i.e., $(\delta X)^{\dagger} = -\delta X$. Hence, we may write
		\begin{equation}
			\label{eq:variation}
			\delta S(U\rho U^{\dagger}||\sigma)=-\tr{\delta X\,\com{U\rho U^\dagger}{\lo{\sigma}}}=0
		\end{equation}
		
		Then, a solution to   Eq.~\eqref{eq:variation} is given by
		\begin{equation}
			\label{eq:Uopt}
			\widetilde{U}\equiv \sumint_i 
			\ket{s_i}\bra{p_i}\,,
		\end{equation}
		for which we immediately obtain
		\begin{equation}
			S(\widetilde{U}\rho \widetilde{U}^\dagger||\sigma)=\sumint_i p_i 
			\lo{p_i/s_i} \equiv D\left(\rho||\sigma\right)\,.
		\end{equation}
		
		Therefore, we conclude that the minimum of the quantum relative entropy under all unitary transformations of $\rho$ is nothing but the classical relative entropy of its distribution of eigenvalues. 
	\end{proof}

	\section{Proof of   Eq.~\eqref{eq:ergo_2}}
	\label{app:ErgotropyRelative}
	\begin{proof}
		In this section, we prove   Eq.~\eqref{eq:ergo_2}. Let $\mc{E}(\rho)$ be the quantum ergotropy of a quantum state
		\begin{equation}
			\rho=\sumint_i p_i \ket{p_i}\bra{p_i}\quad\text{with}\quad  p_i \geq p_{i+1}\,,
		\end{equation}
		and $\rho^{\eq}$ be the Gibbs state
		\begin{equation}
			\rho^{\eq} = \frac{e^{-\beta H}}{Z} = \sumint_j\frac{e^{-\beta E_j}}{Z}\ket{E_j}\bra{E_j}\quad\text{with}\quad E_j\leq E_{j+1}\,
		\end{equation} 
		with an arbitrary inverse temperature $\beta$. Then, 
		{quantum} ergotropy is given by
		\begin{equation}
			\mc{E}(\rho) = \sumint_{i,j}p_iE_j|\braket{E_j}{p_i}|^2-\sumint_{i}p_iE_i\,.
			\label{eq:DefErgo}
		\end{equation}
		
		The quantum relative entropy $S\left(\rho||\rho^{\eq}\right)$ is explicitly written as 
		\begin{equation}
			\begin{split}
				S\left(\rho||\rho^{\eq}\right) &= \tr{\rho\lo{\rho}}+\beta\tr{\rho H}+\lo{Z}\\
				&=\sumint_{i}p_i\lo{p_i}+\beta \sumint_{i,j}p_iE_j|\braket{E_j}{p_i}|^2+\lo{Z}\,,
			\end{split}
			\label{eq:QR}
		\end{equation}
		and the classical relative entropy with respect to the eigenvalue distributions is
		\begin{equation}
			D\left(\rho||\rho^{\eq}\right) = \sumint_{i}p_i\lo{p_i} + \beta\sumint_{i}p_iE_i+\lo{Z}\,.
			\label{eq:CR}
		\end{equation}
		
		Therefore, we can obtain   Eq.~\eqref{eq:ergo_2} 
		\begin{align}
			\beta\mc{E}(\rho) =  S\left(\rho||\rho^{\eq}\right)-D\left(\rho||\rho^{\eq}\right)\,.
		\end{align}
		
		Note that $\beta^{-1}\left(S\left(\rho||\rho^{\eq}\right)-D\left(\rho||\rho^{\eq}\right)\right)$ makes $\mc{E}(\rho)$ independent of $\beta$. 
	\end{proof}

	\section{{Non-Negativity of Divergence-Like Quantity in \mbox{Thermodynamic Scenario}}}
	\label{app:divergence-like}
	{In this section, let us explain why the divergence-like quantity introduced in \mbox{  Eq.~\eqref{eq:relative}} takes only non-negative values in the thermodynamic scenario, while in general, $i$ could take negative values. }
	
	Consider the relative entropy of the probability distribution
	\begin{equation}
		p_{B|A}(\Gamma',\Gamma)=p(\Gamma'|\Gamma)p_A(\Gamma)
	\end{equation}
	with respect to a certain probability distribution $r_B(\Gamma')$. Let us write
	\begin{equation}
		p_B(\Gamma') \equiv \int d\Gamma p_{B|A}(\Gamma',\Gamma)=\int d\Gamma p(\Gamma'|\Gamma)p_A(\Gamma)\,.
	\end{equation}
	
	Then, we have
	\begin{equation}
		\label{appeq:divergence-like}
		\begin{split}
			D\left(p_{B|A}||r_B\right) &= \int d\Gamma' d\Gamma p_{B|A}(\Gamma',\Gamma)\lo{\frac{p_{B|A}(\Gamma',\Gamma)}{r_B(\Gamma')}}\\
			&=\int d\Gamma' d\Gamma p_{B|A}(\Gamma',\Gamma)\lo{ p_{B|A}(\Gamma',\Gamma)}-\int d\Gamma' p_{B}(\Gamma')\lo{r_B(\Gamma')}\,.
		\end{split}
	\end{equation}
	
	Because of the vanishing conditional entropy due to the Liouville's equation,
	\begin{equation}
		\mc{H}(B|A)\equiv-\int d\Gamma'd\Gamma p_{B|A}(\Gamma',\Gamma)\lo{p(\Gamma'|\Gamma)} = 0\,,
	\end{equation}
	and the normalization of the conditional probability distribution in   Eq.~\eqref{eq:norm}, we have
	\begin{equation}
		\int d\Gamma' d\Gamma p_{B|A}(\Gamma',\Gamma)\lo{ p_{B|A}(\Gamma',\Gamma)}=\int d\Gamma p_A(\Gamma)\lo{ p_A(\Gamma)}\,.
	\end{equation}
	
	Therefore, we have
	\begin{equation}
		D\left(p_{B|A}||r_B\right) = \int d\Gamma p_A(\Gamma)\lo{p_A(\Gamma)}-\int d\Gamma' p_B(\Gamma')\lo{r_B(\Gamma')}\,.
	\end{equation}
	
	This could be negative. However, in the thermodynamic scenario, from the second law of thermodynamics, the final entropy $\mc{H}(B)$ has to be greater than or equal to the initial entropy $\mc{H}(A)$ 
	\begin{equation}
		\mc{H}(A)\leq \mc{H}(B) \Longleftrightarrow \int d\Gamma~ p_A(\Gamma)\lo{p_A(\Gamma)} \geq \int d\Gamma'~  p_B(\Gamma')\lo{p_B(\Gamma')}\,.
	\end{equation}
	
	Hence, we have
	\begin{equation}
		D\left(p_{B|A}||r_B\right) \geq \int d\Gamma'~  p_B(\Gamma')\lo{\frac{p_B(\Gamma')}{r_B(\Gamma')}}=
		D(p_B||r_B)\geq 0\,.
	\end{equation}
	
	Therefore, the divergence-like quantity in   Eq.~\eqref{appeq:divergence-like} becomes non-negative
	\begin{equation}
		D\left(p_{B|A}||r_B\right) \geq 0\,.
	\end{equation}

	\section{{Proof of   Eq.~\eqref{eq:MinClassicalRelativeEntropy}}}
	\label{app:MinClassicalRelativeEntropy}
	\begin{proof}
		In this section, we provide a proof for   Eq.~\eqref{eq:MinClassicalRelativeEntropy}. A variation in $\xi$ can be written as
		\begin{equation}
			\delta\xi\equiv\delta\Gamma'\cdot\nabla_{\Gamma'}\xi+\delta\Gamma\cdot\nabla_{\Gamma}\xi\,,
		\end{equation}
		where we replace $\Gamma''$ with $\Gamma'$ without loss of generality. From the expression of $p_B^{\eq}$ and the vanishing conditional entropy due to the Liouvillian evolution, we obtain
		\begin{equation}
			\delta D\left(\mathscr{Q}_{B|A}||p_B^{\eq}\right)\!=\!\beta \int d\Gamma \int d\Gamma' p_A(\Gamma)E_B(\Gamma')\,\delta\xi\,,
		\end{equation}
		where we use the explicit expression for $p_B^{\eq}$. We can find that the variation of the relative entropy vanishes,  $\delta D\left(\mathscr{Q}_{B|A}||p_B^{\eq}\right)=0$, for $\xi(\Gamma' |\Gamma)=\delta(\Gamma' -\Gamma)$; therefore, we obtain
		\begin{equation}
			\min_{\xi}\left[D\left(\mathscr{Q}_{B|A}||p_B^{\eq}\right)\right] = D\left(p_A||p_B^{\eq}\right)\,.
		\end{equation}
	\end{proof}

	\bibliography{refGeo.bib}

\begin{thebibliography}{55}%
\makeatletter
\providecommand \@ifxundefined [1]{%
 \@ifx{#1\undefined}
}%
\providecommand \@ifnum [1]{%
 \ifnum #1\expandafter \@firstoftwo
 \else \expandafter \@secondoftwo
 \fi
}%
\providecommand \@ifx [1]{%
 \ifx #1\expandafter \@firstoftwo
 \else \expandafter \@secondoftwo
 \fi
}%
\providecommand \natexlab [1]{#1}%
\providecommand \enquote  [1]{``#1''}%
\providecommand \bibnamefont  [1]{#1}%
\providecommand \bibfnamefont [1]{#1}%
\providecommand \citenamefont [1]{#1}%
\providecommand \href@noop [0]{\@secondoftwo}%
\providecommand \href [0]{\begingroup \@sanitize@url \@href}%
\providecommand \@href[1]{\@@startlink{#1}\@@href}%
\providecommand \@@href[1]{\endgroup#1\@@endlink}%
\providecommand \@sanitize@url [0]{\catcode `\\12\catcode `\$12\catcode
  `\&12\catcode `\#12\catcode `\^12\catcode `\_12\catcode `\%12\relax}%
\providecommand \@@startlink[1]{}%
\providecommand \@@endlink[0]{}%
\providecommand \url  [0]{\begingroup\@sanitize@url \@url }%
\providecommand \@url [1]{\endgroup\@href {#1}{\urlprefix }}%
\providecommand \urlprefix  [0]{URL }%
\providecommand \Eprint [0]{\href }%
\providecommand \doibase [0]{http://dx.doi.org/}%
\providecommand \selectlanguage [0]{\@gobble}%
\providecommand \bibinfo  [0]{\@secondoftwo}%
\providecommand \bibfield  [0]{\@secondoftwo}%
\providecommand \translation [1]{[#1]}%
\providecommand \BibitemOpen [0]{}%
\providecommand \bibitemStop [0]{}%
\providecommand \bibitemNoStop [0]{.\EOS\space}%
\providecommand \EOS [0]{\spacefactor3000\relax}%
\providecommand \BibitemShut  [1]{\csname bibitem#1\endcsname}%
\let\auto@bib@innerbib\@empty
%</preamble>
\bibitem [{\citenamefont {Gellhorn}(1970)}]{Gellhorn1970}%
  \BibitemOpen
  \bibfield  {author} {\bibinfo {author} {\bibfnamefont {E.}~\bibnamefont
  {Gellhorn}},\ }\bibfield  {title} {\enquote {\bibinfo {title} {The emotions
  and the ergotropic and trophotropic systems},}\ }\href {\doibase
  10.1007/BF00422863} {\bibfield  {journal} {\bibinfo  {journal} {Psychol.
  Forsch.}\ }\textbf {\bibinfo {volume} {34}},\ \bibinfo {pages} {67--94}
  (\bibinfo {year} {1970})}\BibitemShut {NoStop}%
\bibitem [{\citenamefont {Allahverdyan}\ \emph {et~al.}(2004)\citenamefont
  {Allahverdyan}, \citenamefont {Balian},\ and\ \citenamefont
  {Nieuwenhuizen}}]{Allahverdyan04}%
  \BibitemOpen
  \bibfield  {author} {\bibinfo {author} {\bibfnamefont {A.~E.}\ \bibnamefont
  {Allahverdyan}}, \bibinfo {author} {\bibfnamefont {R.}~\bibnamefont
  {Balian}}, \ and\ \bibinfo {author} {\bibfnamefont {Th.~M.}\ \bibnamefont
  {Nieuwenhuizen}},\ }\bibfield  {title} {\enquote {\bibinfo {title} {Maximal
  work extraction from finite quantum systems},}\ }\href
  {https://iopscience.iop.org/article/10.1209/epl/i2004-10101-2} {\bibfield
  {journal} {\bibinfo  {journal} {EPL (Europhys. Lett.)}\ }\textbf {\bibinfo
  {volume} {67}},\ \bibinfo {pages} {565} (\bibinfo {year} {2004})}\BibitemShut
  {NoStop}%
\bibitem [{\citenamefont {Pusz}\ and\ \citenamefont
  {Woronowicz}(1978)}]{Pusz1978}%
  \BibitemOpen
  \bibfield  {author} {\bibinfo {author} {\bibfnamefont {W.}~\bibnamefont
  {Pusz}}\ and\ \bibinfo {author} {\bibfnamefont {S.~L.}\ \bibnamefont
  {Woronowicz}},\ }\bibfield  {title} {\enquote {\bibinfo {title} {Passive
  states and {KMS} states for general quantum systems},}\ }\href {\doibase
  10.1007/BF01614224} {\bibfield  {journal} {\bibinfo  {journal} {Commun. Math.
  Phys.}\ }\textbf {\bibinfo {volume} {58}},\ \bibinfo {pages} {273--290}
  (\bibinfo {year} {1978})}\BibitemShut {NoStop}%
\bibitem [{\citenamefont {Koukoulekidis}\ \emph {et~al.}(2021)\citenamefont
  {Koukoulekidis}, \citenamefont {Alexander}, \citenamefont {Hebdige},\ and\
  \citenamefont {Jennings}}]{Koukoulekidis19}%
  \BibitemOpen
  \bibfield  {author} {\bibinfo {author} {\bibfnamefont {Nikolaos}\
  \bibnamefont {Koukoulekidis}}, \bibinfo {author} {\bibfnamefont {Rhea}\
  \bibnamefont {Alexander}}, \bibinfo {author} {\bibfnamefont {Thomas}\
  \bibnamefont {Hebdige}}, \ and\ \bibinfo {author} {\bibfnamefont {David}\
  \bibnamefont {Jennings}},\ }\bibfield  {title} {\enquote {\bibinfo {title}
  {The geometry of passivity for quantum systems and a novel elementary
  derivation of the {G}ibbs state},}\ }\href
  {https://quantum-journal.org/papers/q-2021-03-15-411/} {\bibfield  {journal}
  {\bibinfo  {journal} {Quantum}\ }\textbf {\bibinfo {volume} {5}},\ \bibinfo
  {pages} {411} (\bibinfo {year} {2021})}\BibitemShut {NoStop}%
\bibitem [{\citenamefont {G{\'{o}}recki}\ and\ \citenamefont
  {Pusz}(1980)}]{Gorecki80}%
  \BibitemOpen
  \bibfield  {author} {\bibinfo {author} {\bibfnamefont {J.}~\bibnamefont
  {G{\'{o}}recki}}\ and\ \bibinfo {author} {\bibfnamefont {W.}~\bibnamefont
  {Pusz}},\ }\bibfield  {title} {\enquote {\bibinfo {title} {Passive states for
  finite classical systems},}\ }\href {\doibase
  https://doi.org/10.1007/BF00943428} {\bibfield  {journal} {\bibinfo
  {journal} {Lett. Math. Phys.}\ }\textbf {\bibinfo {volume} {4}},\ \bibinfo
  {pages} {433} (\bibinfo {year} {1980})}\BibitemShut {NoStop}%
\bibitem [{\citenamefont {Dani{\"{e}}ls}(1981)}]{Daniels81}%
  \BibitemOpen
  \bibfield  {author} {\bibinfo {author} {\bibfnamefont {H.~A.~M.}\
  \bibnamefont {Dani{\"{e}}ls}},\ }\bibfield  {title} {\enquote {\bibinfo
  {title} {Passivity and equilibrium for classical hamiltonian systems},}\
  }\href {\doibase https://doi.org/10.1063/1.524949} {\bibfield  {journal}
  {\bibinfo  {journal} {J. Math. Phys.}\ }\textbf {\bibinfo {volume} {22}},\
  \bibinfo {pages} {843} (\bibinfo {year} {1981})}\BibitemShut {NoStop}%
\bibitem [{\citenamefont {Deffner}\ and\ \citenamefont
  {Campbell}(2019)}]{DeffnerBook19}%
  \BibitemOpen
  \bibfield  {author} {\bibinfo {author} {\bibfnamefont {S.}~\bibnamefont
  {Deffner}}\ and\ \bibinfo {author} {\bibfnamefont {S.}~\bibnamefont
  {Campbell}},\ }\href {https://iopscience.iop.org/book/978-1-64327-658-8}
  {\emph {\bibinfo {title} {Quantum Thermodynamics}}}\ (\bibinfo  {publisher}
  {Morgan and Claypool Publishers, San Rafael},\ \bibinfo {year}
  {2019})\BibitemShut {NoStop}%
\bibitem [{\citenamefont {Goold}\ \emph {et~al.}(2016)\citenamefont {Goold},
  \citenamefont {Huber}, \citenamefont {Riera}, \citenamefont {del Rio},\ and\
  \citenamefont {Skrzypczyk}}]{Goold16}%
  \BibitemOpen
  \bibfield  {author} {\bibinfo {author} {\bibfnamefont {John}\ \bibnamefont
  {Goold}}, \bibinfo {author} {\bibfnamefont {Marcus}\ \bibnamefont {Huber}},
  \bibinfo {author} {\bibfnamefont {Arnau}\ \bibnamefont {Riera}}, \bibinfo
  {author} {\bibfnamefont {L{\'{i}}dia}\ \bibnamefont {del Rio}}, \ and\
  \bibinfo {author} {\bibfnamefont {Paul}\ \bibnamefont {Skrzypczyk}},\
  }\bibfield  {title} {\enquote {\bibinfo {title} {The role of quantum
  information in thermodynamics—a topical review},}\ }\href
  {https://iopscience.iop.org/article/10.1088/1751-8113/49/14/143001/meta}
  {\bibfield  {journal} {\bibinfo  {journal} {J. Phys. A: Math. Theor.}\
  }\textbf {\bibinfo {volume} {49}},\ \bibinfo {pages} {143001} (\bibinfo
  {year} {2016})}\BibitemShut {NoStop}%
\bibitem [{\citenamefont {Perarnau-Llobet}\ \emph {et~al.}(2017)\citenamefont
  {Perarnau-Llobet}, \citenamefont {B\"{a}umer}, \citenamefont {Hovhannisyan},
  \citenamefont {Huber},\ and\ \citenamefont {Acin}}]{Marti17}%
  \BibitemOpen
  \bibfield  {author} {\bibinfo {author} {\bibfnamefont {Mart\'{l}}\
  \bibnamefont {Perarnau-Llobet}}, \bibinfo {author} {\bibfnamefont {Elisa}\
  \bibnamefont {B\"{a}umer}}, \bibinfo {author} {\bibfnamefont {Karen~V.}\
  \bibnamefont {Hovhannisyan}}, \bibinfo {author} {\bibfnamefont {Marcus}\
  \bibnamefont {Huber}}, \ and\ \bibinfo {author} {\bibfnamefont {Antonio}\
  \bibnamefont {Acin}},\ }\bibfield  {title} {\enquote {\bibinfo {title} {No-go
  theorem for the characterization of work fluctuations in coherent quantum
  systems},}\ }\href
  {https://journals.aps.org/prl/abstract/10.1103/PhysRevLett.118.070601}
  {\bibfield  {journal} {\bibinfo  {journal} {Phys. Rev. Lett.}\ }\textbf
  {\bibinfo {volume} {118}},\ \bibinfo {pages} {070601} (\bibinfo {year}
  {2017})}\BibitemShut {NoStop}%
\bibitem [{\citenamefont {Levy}\ and\ \citenamefont
  {Lostaglio}(2020)}]{Levy19}%
  \BibitemOpen
  \bibfield  {author} {\bibinfo {author} {\bibfnamefont {Amikam}\ \bibnamefont
  {Levy}}\ and\ \bibinfo {author} {\bibfnamefont {Matteo}\ \bibnamefont
  {Lostaglio}},\ }\bibfield  {title} {\enquote {\bibinfo {title} {A
  quasiprobability distribution for heat fluctuations in the quantum regime},}\
  }\href {\doibase 10.1103/PRXQuantum.1.010309} {\bibfield  {journal} {\bibinfo
   {journal} {Phys. Rev. X Quantum}\ }\textbf {\bibinfo {volume} {1}},\
  \bibinfo {pages} {010309} (\bibinfo {year} {2020})}\BibitemShut {NoStop}%
\bibitem [{\citenamefont {Santos}\ \emph {et~al.}(2019)\citenamefont {Santos},
  \citenamefont {C{\'{e}}leri}, \citenamefont {Landi},\ and\ \citenamefont
  {Paternostro}}]{Santos19}%
  \BibitemOpen
  \bibfield  {author} {\bibinfo {author} {\bibfnamefont {Jader~P.}\
  \bibnamefont {Santos}}, \bibinfo {author} {\bibfnamefont {Lucas~C.}\
  \bibnamefont {C{\'{e}}leri}}, \bibinfo {author} {\bibfnamefont {Gabriel~T.}\
  \bibnamefont {Landi}}, \ and\ \bibinfo {author} {\bibfnamefont {Mauro}\
  \bibnamefont {Paternostro}},\ }\bibfield  {title} {\enquote {\bibinfo {title}
  {The role of quantum coherence in non-equilibrium entropy production},}\
  }\href {\doibase 10.1038/s41534-019-0138-y} {\bibfield  {journal} {\bibinfo
  {journal} {npj Quantum Inf.}\ }\textbf {\bibinfo {volume} {5}},\ \bibinfo
  {pages} {23} (\bibinfo {year} {2019})}\BibitemShut {NoStop}%
\bibitem [{\citenamefont {Niedenzu}\ \emph {et~al.}(2018)\citenamefont
  {Niedenzu}, \citenamefont {Mukherjee}, \citenamefont {Ghosh}, \citenamefont
  {Kofman},\ and\ \citenamefont {Kurizki}}]{Niedenzu2018}%
  \BibitemOpen
  \bibfield  {author} {\bibinfo {author} {\bibfnamefont {Wolfgang}\
  \bibnamefont {Niedenzu}}, \bibinfo {author} {\bibfnamefont {Victor}\
  \bibnamefont {Mukherjee}}, \bibinfo {author} {\bibfnamefont {Arnab}\
  \bibnamefont {Ghosh}}, \bibinfo {author} {\bibfnamefont {Abraham~G.}\
  \bibnamefont {Kofman}}, \ and\ \bibinfo {author} {\bibfnamefont {Gershon}\
  \bibnamefont {Kurizki}},\ }\bibfield  {title} {\enquote {\bibinfo {title}
  {Quantum engine efficiency bound beyond the second law of thermodynamics},}\
  }\href {\doibase 10.1038/s41467-017-01991-6} {\bibfield  {journal} {\bibinfo
  {journal} {Nat. Commun.}\ }\textbf {\bibinfo {volume} {9}},\ \bibinfo {pages}
  {165} (\bibinfo {year} {2018})}\BibitemShut {NoStop}%
\bibitem [{\citenamefont {Cherubim}\ \emph {et~al.}(2019)\citenamefont
  {Cherubim}, \citenamefont {Brito},\ and\ \citenamefont
  {Deffner}}]{Cherubim2019}%
  \BibitemOpen
  \bibfield  {author} {\bibinfo {author} {\bibfnamefont {Cleverson}\
  \bibnamefont {Cherubim}}, \bibinfo {author} {\bibfnamefont {Frederico}\
  \bibnamefont {Brito}}, \ and\ \bibinfo {author} {\bibfnamefont {Sebastian}\
  \bibnamefont {Deffner}},\ }\bibfield  {title} {\enquote {\bibinfo {title}
  {Non-thermal quantum engine in transmon qubits},}\ }\href
  {https://www.mdpi.com/1099-4300/21/6/545} {\bibfield  {journal} {\bibinfo
  {journal} {Entropy}\ }\textbf {\bibinfo {volume} {21}},\ \bibinfo {pages}
  {545} (\bibinfo {year} {2019})}\BibitemShut {NoStop}%
\bibitem [{\citenamefont {Francica}\ \emph {et~al.}(2020)\citenamefont
  {Francica}, \citenamefont {Binder}, \citenamefont {Guarnieri}, \citenamefont
  {Mitchison}, \citenamefont {Goold},\ and\ \citenamefont
  {Plastina}}]{Francica20}%
  \BibitemOpen
  \bibfield  {author} {\bibinfo {author} {\bibfnamefont {G.}~\bibnamefont
  {Francica}}, \bibinfo {author} {\bibfnamefont {F.C.}\ \bibnamefont {Binder}},
  \bibinfo {author} {\bibfnamefont {G.}~\bibnamefont {Guarnieri}}, \bibinfo
  {author} {\bibfnamefont {M.T.}\ \bibnamefont {Mitchison}}, \bibinfo {author}
  {\bibfnamefont {J.}~\bibnamefont {Goold}}, \ and\ \bibinfo {author}
  {\bibfnamefont {F.}~\bibnamefont {Plastina}},\ }\bibfield  {title} {\enquote
  {\bibinfo {title} {Quantum coherence and ergotropy},}\ }\href
  {https://journals.aps.org/prl/abstract/10.1103/PhysRevLett.125.180603}
  {\bibfield  {journal} {\bibinfo  {journal} {Phys. Rev. Lett.}\ }\textbf
  {\bibinfo {volume} {125}},\ \bibinfo {pages} {180603} (\bibinfo {year}
  {2020})}\BibitemShut {NoStop}%
\bibitem [{\citenamefont {\ifmmode~\mbox{\c{C}}\else
  \c{C}\fi{}akmak}(2020)}]{Cakmak2020}%
  \BibitemOpen
  \bibfield  {author} {\bibinfo {author} {\bibfnamefont {B.}~\bibnamefont
  {\ifmmode~\mbox{\c{C}}\else \c{C}\fi{}akmak}},\ }\bibfield  {title} {\enquote
  {\bibinfo {title} {Ergotropy from coherences in an open quantum system},}\
  }\href {\doibase 10.1103/PhysRevE.102.042111} {\bibfield  {journal} {\bibinfo
   {journal} {Phys. Rev. E}\ }\textbf {\bibinfo {volume} {102}},\ \bibinfo
  {pages} {042111} (\bibinfo {year} {2020})}\BibitemShut {NoStop}%
\bibitem [{\citenamefont {Francica}\ \emph {et~al.}(2017)\citenamefont
  {Francica}, \citenamefont {Goold}, \citenamefont {Plastina},\ and\
  \citenamefont {Paternostro}}]{Francica2017}%
  \BibitemOpen
  \bibfield  {author} {\bibinfo {author} {\bibfnamefont {Gianluca}\
  \bibnamefont {Francica}}, \bibinfo {author} {\bibfnamefont {John}\
  \bibnamefont {Goold}}, \bibinfo {author} {\bibfnamefont {Francesco}\
  \bibnamefont {Plastina}}, \ and\ \bibinfo {author} {\bibfnamefont {Mauro}\
  \bibnamefont {Paternostro}},\ }\bibfield  {title} {\enquote {\bibinfo {title}
  {Daemonic ergotropy: enhanced work extraction from quantum correlations},}\
  }\href {\doibase 10.1038/s41534-017-0012-8} {\bibfield  {journal} {\bibinfo
  {journal} {npj Quantum Inf.}\ }\textbf {\bibinfo {volume} {3}},\ \bibinfo
  {pages} {12} (\bibinfo {year} {2017})}\BibitemShut {NoStop}%
\bibitem [{\citenamefont {Touil}\ \emph {et~al.}(2021)\citenamefont {Touil},
  \citenamefont {\c{C}akmak},\ and\ \citenamefont {Deffner}}]{Touil2021}%
  \BibitemOpen
  \bibfield  {author} {\bibinfo {author} {\bibfnamefont {Akram}\ \bibnamefont
  {Touil}}, \bibinfo {author} {\bibfnamefont {Bar\i\c{s}}\ \bibnamefont
  {\c{C}akmak}}, \ and\ \bibinfo {author} {\bibfnamefont {Sebastian}\
  \bibnamefont {Deffner}},\ }\bibfield  {title} {\enquote {\bibinfo {title}
  {Second law of thermodynamics for quantum correlations},}\ }\href
  {https://arxiv.org/abs/2102.13606} {\bibfield  {journal} {\bibinfo  {journal}
  {arXiv preprint arXiv:2102.13606}\ } (\bibinfo {year} {2021})}\BibitemShut
  {NoStop}%
\bibitem [{\citenamefont {Smith}(2019)}]{Smith19Thesis}%
  \BibitemOpen
  \bibfield  {author} {\bibinfo {author} {\bibfnamefont {Andrew~Maven}\
  \bibnamefont {Smith}},\ }\emph {\bibinfo {title} {Studies in Nonequilibrium
  Quantum Thermodynamics}},\ \href {https://drum.lib.umd.edu/handle/1903/26061}
  {Ph.D. thesis},\ \bibinfo  {school} {University of Maryland, College Park}
  (\bibinfo {year} {2019})\BibitemShut {NoStop}%
\bibitem [{\citenamefont {Smith}\ \emph {et~al.}()\citenamefont {Smith},
  \citenamefont {Sinha},\ and\ \citenamefont {Jarzynski}}]{Smith2021}%
  \BibitemOpen
  \bibfield  {author} {\bibinfo {author} {\bibfnamefont {A.}~\bibnamefont
  {Smith}}, \bibinfo {author} {\bibfnamefont {K.}~\bibnamefont {Sinha}}, \ and\
  \bibinfo {author} {\bibfnamefont {C.}~\bibnamefont {Jarzynski}},\ }\href@noop
  {} {\ }\bibinfo {note} {(to be published)}\BibitemShut {NoStop}%
\bibitem [{\citenamefont {Anza}\ and\ \citenamefont
  {Crutchfield}(2020{\natexlab{a}})}]{Anza20a}%
  \BibitemOpen
  \bibfield  {author} {\bibinfo {author} {\bibfnamefont {Fabio}\ \bibnamefont
  {Anza}}\ and\ \bibinfo {author} {\bibfnamefont {James~P.}\ \bibnamefont
  {Crutchfield}},\ }\bibfield  {title} {\enquote {\bibinfo {title} {Geometric
  quantum state estimation},}\ }\href {https://arxiv.org/abs/2008.08679}
  {\bibfield  {journal} {\bibinfo  {journal} {arXiv:2008.08679}\ } (\bibinfo
  {year} {2020}{\natexlab{a}})}\BibitemShut {NoStop}%
\bibitem [{\citenamefont {Anza}\ and\ \citenamefont
  {Crutchfield}(2020{\natexlab{b}})}]{Anza20b}%
  \BibitemOpen
  \bibfield  {author} {\bibinfo {author} {\bibfnamefont {Fabio}\ \bibnamefont
  {Anza}}\ and\ \bibinfo {author} {\bibfnamefont {James~P.}\ \bibnamefont
  {Crutchfield}},\ }\bibfield  {title} {\enquote {\bibinfo {title} {Beyond
  density matrices: Geometric quantum states},}\ }\href
  {https://arxiv.org/abs/2008.08682} {\bibfield  {journal} {\bibinfo  {journal}
  {arXiv:2008.08682}\ } (\bibinfo {year} {2020}{\natexlab{b}})}\BibitemShut
  {NoStop}%
\bibitem [{\citenamefont {Anza}\ and\ \citenamefont
  {Crutchfield}(2020{\natexlab{c}})}]{Anza20c}%
  \BibitemOpen
  \bibfield  {author} {\bibinfo {author} {\bibfnamefont {Fabio}\ \bibnamefont
  {Anza}}\ and\ \bibinfo {author} {\bibfnamefont {James~P.}\ \bibnamefont
  {Crutchfield}},\ }\bibfield  {title} {\enquote {\bibinfo {title} {Geometric
  quantum thermodynamics},}\ }\href {https://arxiv.org/abs/2008.08683}
  {\bibfield  {journal} {\bibinfo  {journal} {arXiv:2008.08683}\ } (\bibinfo
  {year} {2020}{\natexlab{c}})}\BibitemShut {NoStop}%
\bibitem [{\citenamefont {Deffner}\ \emph {et~al.}(2016)\citenamefont
  {Deffner}, \citenamefont {Paz},\ and\ \citenamefont {Zurek}}]{Deffner16}%
  \BibitemOpen
  \bibfield  {author} {\bibinfo {author} {\bibfnamefont {Sebastian}\
  \bibnamefont {Deffner}}, \bibinfo {author} {\bibfnamefont {Juan~Pablo}\
  \bibnamefont {Paz}}, \ and\ \bibinfo {author} {\bibfnamefont {Wojciech~H.}\
  \bibnamefont {Zurek}},\ }\bibfield  {title} {\enquote {\bibinfo {title}
  {Quantum work and the thermodynamic cost of quantum measurements},}\ }\href
  {https://journals.aps.org/pre/abstract/10.1103/PhysRevE.94.010103} {\bibfield
   {journal} {\bibinfo  {journal} {Phys. Rev. E}\ }\textbf {\bibinfo {volume}
  {94}},\ \bibinfo {pages} {010103(R)} (\bibinfo {year} {2016})}\BibitemShut
  {NoStop}%
\bibitem [{\citenamefont {Beyer}\ \emph {et~al.}(2020)\citenamefont {Beyer},
  \citenamefont {Luoma},\ and\ \citenamefont {Strunz}}]{Beyer2020}%
  \BibitemOpen
  \bibfield  {author} {\bibinfo {author} {\bibfnamefont {Konstantin}\
  \bibnamefont {Beyer}}, \bibinfo {author} {\bibfnamefont {Kimmo}\ \bibnamefont
  {Luoma}}, \ and\ \bibinfo {author} {\bibfnamefont {Walter~T.}\ \bibnamefont
  {Strunz}},\ }\bibfield  {title} {\enquote {\bibinfo {title} {Work as an
  external quantum observable and an operational quantum work fluctuation
  theorem},}\ }\href
  {https://link.aps.org/doi/10.1103/PhysRevResearch.2.033508} {\bibfield
  {journal} {\bibinfo  {journal} {Phys. Rev. Research}\ }\textbf {\bibinfo
  {volume} {2}},\ \bibinfo {pages} {033508} (\bibinfo {year}
  {2020})}\BibitemShut {NoStop}%
\bibitem [{\citenamefont {Sone}\ \emph {et~al.}(2020)\citenamefont {Sone},
  \citenamefont {Liu},\ and\ \citenamefont {Cappellaro}}]{Sone20a}%
  \BibitemOpen
  \bibfield  {author} {\bibinfo {author} {\bibfnamefont {Akira}\ \bibnamefont
  {Sone}}, \bibinfo {author} {\bibfnamefont {Yi-Xiang}\ \bibnamefont {Liu}}, \
  and\ \bibinfo {author} {\bibfnamefont {Paola}\ \bibnamefont {Cappellaro}},\
  }\bibfield  {title} {\enquote {\bibinfo {title} {Quantum {J}arzynski equality
  in open quantum systems from the one-time measurement scheme},}\ }\href
  {\doibase 10.1103/PhysRevLett.125.060602} {\bibfield  {journal} {\bibinfo
  {journal} {Phys. Rev. Lett.}\ }\textbf {\bibinfo {volume} {125}},\ \bibinfo
  {pages} {060602} (\bibinfo {year} {2020})}\BibitemShut {NoStop}%
\bibitem [{\citenamefont {Sone}\ and\ \citenamefont {Deffner}(2021)}]{Sone21b}%
  \BibitemOpen
  \bibfield  {author} {\bibinfo {author} {\bibfnamefont {Akira}\ \bibnamefont
  {Sone}}\ and\ \bibinfo {author} {\bibfnamefont {Sebastian}\ \bibnamefont
  {Deffner}},\ }\bibfield  {title} {\enquote {\bibinfo {title} {{J}arzynski
  equality for stochastic conditional work},}\ }\href {\doibase
  10.1007/s10955-021-02720-6} {\bibfield  {journal} {\bibinfo  {journal} {J.
  Stat. Phys}\ }\textbf {\bibinfo {volume} {183}},\ \bibinfo {pages} {11}
  (\bibinfo {year} {2021})}\BibitemShut {NoStop}%
\bibitem [{\citenamefont {Allahverdyan}\ and\ \citenamefont
  {Nieuwenhuizen}(2005)}]{Allahverdyan05}%
  \BibitemOpen
  \bibfield  {author} {\bibinfo {author} {\bibfnamefont {A.~E.}\ \bibnamefont
  {Allahverdyan}}\ and\ \bibinfo {author} {\bibfnamefont {Th.~M.}\ \bibnamefont
  {Nieuwenhuizen}},\ }\bibfield  {title} {\enquote {\bibinfo {title}
  {Fluctuations of work from quantum subensembles: The case against quantum
  work-fluctuation theorems},}\ }\href {\doibase
  https://doi.org/10.1103/PhysRevE.71.066102} {\bibfield  {journal} {\bibinfo
  {journal} {Phys. Rev. E}\ }\textbf {\bibinfo {volume} {71}},\ \bibinfo
  {pages} {066102} (\bibinfo {year} {2005})}\BibitemShut {NoStop}%
\bibitem [{\citenamefont {Kurchan}(2001)}]{Kurchan01}%
  \BibitemOpen
  \bibfield  {author} {\bibinfo {author} {\bibfnamefont {Jorge}\ \bibnamefont
  {Kurchan}},\ }\bibfield  {title} {\enquote {\bibinfo {title} {A quantum
  fluctuation theorem},}\ }\href {https://arxiv.org/abs/cond-mat/0007360v2}
  {\bibfield  {journal} {\bibinfo  {journal} {arXiv:cond-mat/0007360}\ }
  (\bibinfo {year} {2001})}\BibitemShut {NoStop}%
\bibitem [{\citenamefont {Tasaki}(2000)}]{Tasaki00}%
  \BibitemOpen
  \bibfield  {author} {\bibinfo {author} {\bibfnamefont {Hal}\ \bibnamefont
  {Tasaki}},\ }\bibfield  {title} {\enquote {\bibinfo {title} {{J}arzynski
  relations for quantum systems and some applications},}\ }\href
  {https://arxiv.org/abs/cond-mat/0009244v2} {\bibfield  {journal} {\bibinfo
  {journal} {arXiv:cond-mat/0009244}\ } (\bibinfo {year} {2000})}\BibitemShut
  {NoStop}%
\bibitem [{\citenamefont {Talkner}\ \emph {et~al.}(2007)\citenamefont
  {Talkner}, \citenamefont {Lutz},\ and\ \citenamefont
  {H{\"{a}}nggi}}]{Talkner07}%
  \BibitemOpen
  \bibfield  {author} {\bibinfo {author} {\bibfnamefont {Peter}\ \bibnamefont
  {Talkner}}, \bibinfo {author} {\bibfnamefont {Eric}\ \bibnamefont {Lutz}}, \
  and\ \bibinfo {author} {\bibfnamefont {Peter}\ \bibnamefont {H{\"{a}}nggi}},\
  }\bibfield  {title} {\enquote {\bibinfo {title} {Fluctuation theorems: Work
  is not an observable},}\ }\href
  {https://journals.aps.org/pre/abstract/10.1103/PhysRevE.75.050102} {\bibfield
   {journal} {\bibinfo  {journal} {Phys. Rev. E}\ }\textbf {\bibinfo {volume}
  {75}},\ \bibinfo {pages} {050102(R)} (\bibinfo {year} {2007})}\BibitemShut
  {NoStop}%
\bibitem [{\citenamefont {Huber}\ \emph {et~al.}(2008)\citenamefont {Huber},
  \citenamefont {Schmidt-Kaler}, \citenamefont {Deffner},\ and\ \citenamefont
  {Lutz}}]{Huber2008}%
  \BibitemOpen
  \bibfield  {author} {\bibinfo {author} {\bibfnamefont {Gerhard}\ \bibnamefont
  {Huber}}, \bibinfo {author} {\bibfnamefont {Ferdinand}\ \bibnamefont
  {Schmidt-Kaler}}, \bibinfo {author} {\bibfnamefont {Sebastian}\ \bibnamefont
  {Deffner}}, \ and\ \bibinfo {author} {\bibfnamefont {Eric}\ \bibnamefont
  {Lutz}},\ }\bibfield  {title} {\enquote {\bibinfo {title} {Employing trapped
  cold ions to verify the quantum jarzynski equality},}\ }\href {\doibase
  10.1103/PhysRevLett.101.070403} {\bibfield  {journal} {\bibinfo  {journal}
  {Phys. Rev. Lett.}\ }\textbf {\bibinfo {volume} {101}},\ \bibinfo {pages}
  {070403} (\bibinfo {year} {2008})}\BibitemShut {NoStop}%
\bibitem [{\citenamefont {Campisi}\ \emph {et~al.}(2011)\citenamefont
  {Campisi}, \citenamefont {H\"{a}nggi},\ and\ \citenamefont
  {Talkner}}]{Campisi11}%
  \BibitemOpen
  \bibfield  {author} {\bibinfo {author} {\bibfnamefont {Michele}\ \bibnamefont
  {Campisi}}, \bibinfo {author} {\bibfnamefont {Peter}\ \bibnamefont
  {H\"{a}nggi}}, \ and\ \bibinfo {author} {\bibfnamefont {Peter}\ \bibnamefont
  {Talkner}},\ }\bibfield  {title} {\enquote {\bibinfo {title} {Colloquium:
  Quantum fluctuation relations: Foundations and applications},}\ }\href
  {https://journals.aps.org/rmp/abstract/10.1103/RevModPhys.83.771} {\bibfield
  {journal} {\bibinfo  {journal} {Rev. Mod. Phys.}\ }\textbf {\bibinfo {volume}
  {83}},\ \bibinfo {pages} {771} (\bibinfo {year} {2011})}\BibitemShut
  {NoStop}%
\bibitem [{\citenamefont {Deffner}\ and\ \citenamefont
  {Lutz}(2011)}]{Deffner2011PRL}%
  \BibitemOpen
  \bibfield  {author} {\bibinfo {author} {\bibfnamefont {Sebastian}\
  \bibnamefont {Deffner}}\ and\ \bibinfo {author} {\bibfnamefont {Eric}\
  \bibnamefont {Lutz}},\ }\bibfield  {title} {\enquote {\bibinfo {title}
  {Nonequilibrium entropy production for open quantum systems},}\ }\href
  {\doibase 10.1103/PhysRevLett.107.140404} {\bibfield  {journal} {\bibinfo
  {journal} {Phys. Rev. Lett.}\ }\textbf {\bibinfo {volume} {107}},\ \bibinfo
  {pages} {140404} (\bibinfo {year} {2011})}\BibitemShut {NoStop}%
\bibitem [{\citenamefont {Kafri}\ and\ \citenamefont
  {Deffner}(2012)}]{Kafri2012}%
  \BibitemOpen
  \bibfield  {author} {\bibinfo {author} {\bibfnamefont {Dvir}\ \bibnamefont
  {Kafri}}\ and\ \bibinfo {author} {\bibfnamefont {Sebastian}\ \bibnamefont
  {Deffner}},\ }\bibfield  {title} {\enquote {\bibinfo {title} {Holevo's bound
  from a general quantum fluctuation theorem},}\ }\href {\doibase
  10.1103/PhysRevA.86.044302} {\bibfield  {journal} {\bibinfo  {journal} {Phys.
  Rev. A}\ }\textbf {\bibinfo {volume} {86}},\ \bibinfo {pages} {044302}
  (\bibinfo {year} {2012})}\BibitemShut {NoStop}%
\bibitem [{\citenamefont {Mazzola}\ \emph {et~al.}(2013)\citenamefont
  {Mazzola}, \citenamefont {De~Chiara},\ and\ \citenamefont
  {Paternostro}}]{Mazzola2013}%
  \BibitemOpen
  \bibfield  {author} {\bibinfo {author} {\bibfnamefont {L.}~\bibnamefont
  {Mazzola}}, \bibinfo {author} {\bibfnamefont {G.}~\bibnamefont {De~Chiara}},
  \ and\ \bibinfo {author} {\bibfnamefont {M.}~\bibnamefont {Paternostro}},\
  }\bibfield  {title} {\enquote {\bibinfo {title} {Measuring the characteristic
  function of the work distribution},}\ }\href {\doibase
  10.1103/PhysRevLett.110.230602} {\bibfield  {journal} {\bibinfo  {journal}
  {Phys. Rev. Lett.}\ }\textbf {\bibinfo {volume} {110}},\ \bibinfo {pages}
  {230602} (\bibinfo {year} {2013})}\BibitemShut {NoStop}%
\bibitem [{\citenamefont {Dorner}\ \emph {et~al.}(2013)\citenamefont {Dorner},
  \citenamefont {Clark}, \citenamefont {Heaney}, \citenamefont {Fazio},
  \citenamefont {Goold},\ and\ \citenamefont {Vedral}}]{Dorner2013}%
  \BibitemOpen
  \bibfield  {author} {\bibinfo {author} {\bibfnamefont {R.}~\bibnamefont
  {Dorner}}, \bibinfo {author} {\bibfnamefont {S.~R.}\ \bibnamefont {Clark}},
  \bibinfo {author} {\bibfnamefont {L.}~\bibnamefont {Heaney}}, \bibinfo
  {author} {\bibfnamefont {R.}~\bibnamefont {Fazio}}, \bibinfo {author}
  {\bibfnamefont {J.}~\bibnamefont {Goold}}, \ and\ \bibinfo {author}
  {\bibfnamefont {V.}~\bibnamefont {Vedral}},\ }\bibfield  {title} {\enquote
  {\bibinfo {title} {Extracting quantum work statistics and fluctuation
  theorems by single-qubit interferometry},}\ }\href {\doibase
  10.1103/PhysRevLett.110.230601} {\bibfield  {journal} {\bibinfo  {journal}
  {Phys. Rev. Lett.}\ }\textbf {\bibinfo {volume} {110}},\ \bibinfo {pages}
  {230601} (\bibinfo {year} {2013})}\BibitemShut {NoStop}%
\bibitem [{\citenamefont {Roncaglia}\ \emph {et~al.}(2014)\citenamefont
  {Roncaglia}, \citenamefont {Cerisola},\ and\ \citenamefont
  {Paz}}]{Roncaglia2014}%
  \BibitemOpen
  \bibfield  {author} {\bibinfo {author} {\bibfnamefont {Augusto~J.}\
  \bibnamefont {Roncaglia}}, \bibinfo {author} {\bibfnamefont {Federico}\
  \bibnamefont {Cerisola}}, \ and\ \bibinfo {author} {\bibfnamefont
  {Juan~Pablo}\ \bibnamefont {Paz}},\ }\bibfield  {title} {\enquote {\bibinfo
  {title} {Work measurement as a generalized quantum measurement},}\ }\href
  {\doibase 10.1103/PhysRevLett.113.250601} {\bibfield  {journal} {\bibinfo
  {journal} {Phys. Rev. Lett.}\ }\textbf {\bibinfo {volume} {113}},\ \bibinfo
  {pages} {250601} (\bibinfo {year} {2014})}\BibitemShut {NoStop}%
\bibitem [{\citenamefont {Batalh\~ao}\ \emph {et~al.}(2014)\citenamefont
  {Batalh\~ao}, \citenamefont {Souza}, \citenamefont {Mazzola}, \citenamefont
  {Auccaise}, \citenamefont {Sarthour}, \citenamefont {Oliveira}, \citenamefont
  {Goold}, \citenamefont {De~Chiara}, \citenamefont {Paternostro},\ and\
  \citenamefont {Serra}}]{Batalhao2014}%
  \BibitemOpen
  \bibfield  {author} {\bibinfo {author} {\bibfnamefont {Tiago~B.}\
  \bibnamefont {Batalh\~ao}}, \bibinfo {author} {\bibfnamefont {Alexandre~M.}\
  \bibnamefont {Souza}}, \bibinfo {author} {\bibfnamefont {Laura}\ \bibnamefont
  {Mazzola}}, \bibinfo {author} {\bibfnamefont {Ruben}\ \bibnamefont
  {Auccaise}}, \bibinfo {author} {\bibfnamefont {Roberto~S.}\ \bibnamefont
  {Sarthour}}, \bibinfo {author} {\bibfnamefont {Ivan~S.}\ \bibnamefont
  {Oliveira}}, \bibinfo {author} {\bibfnamefont {John}\ \bibnamefont {Goold}},
  \bibinfo {author} {\bibfnamefont {Gabriele}\ \bibnamefont {De~Chiara}},
  \bibinfo {author} {\bibfnamefont {Mauro}\ \bibnamefont {Paternostro}}, \ and\
  \bibinfo {author} {\bibfnamefont {Roberto~M.}\ \bibnamefont {Serra}},\
  }\bibfield  {title} {\enquote {\bibinfo {title} {Experimental reconstruction
  of work distribution and study of fluctuation relations in a closed quantum
  system},}\ }\href {\doibase 10.1103/PhysRevLett.113.140601} {\bibfield
  {journal} {\bibinfo  {journal} {Phys. Rev. Lett.}\ }\textbf {\bibinfo
  {volume} {113}},\ \bibinfo {pages} {140601} (\bibinfo {year}
  {2014})}\BibitemShut {NoStop}%
\bibitem [{\citenamefont {An}\ \emph {et~al.}(2015)\citenamefont {An},
  \citenamefont {Zhang}, \citenamefont {Um}, \citenamefont {Lv}, \citenamefont
  {Lu}, \citenamefont {Zhang}, \citenamefont {Yin}, \citenamefont {Quan},\ and\
  \citenamefont {Kim}}]{An2015}%
  \BibitemOpen
  \bibfield  {author} {\bibinfo {author} {\bibfnamefont {Shuoming}\
  \bibnamefont {An}}, \bibinfo {author} {\bibfnamefont {Jing-Ning}\
  \bibnamefont {Zhang}}, \bibinfo {author} {\bibfnamefont {Mark}\ \bibnamefont
  {Um}}, \bibinfo {author} {\bibfnamefont {Dingshun}\ \bibnamefont {Lv}},
  \bibinfo {author} {\bibfnamefont {Yao}\ \bibnamefont {Lu}}, \bibinfo {author}
  {\bibfnamefont {Junhua}\ \bibnamefont {Zhang}}, \bibinfo {author}
  {\bibfnamefont {Zhang-Qi}\ \bibnamefont {Yin}}, \bibinfo {author}
  {\bibfnamefont {H.~T.}\ \bibnamefont {Quan}}, \ and\ \bibinfo {author}
  {\bibfnamefont {Kihwan}\ \bibnamefont {Kim}},\ }\bibfield  {title} {\enquote
  {\bibinfo {title} {Experimental test of the quantum jarzynski equality with a
  trapped-ion system},}\ }\href {\doibase 10.1038/nphys3197} {\bibfield
  {journal} {\bibinfo  {journal} {Nature Physics}\ }\textbf {\bibinfo {volume}
  {11}},\ \bibinfo {pages} {193--199} (\bibinfo {year} {2015})}\BibitemShut
  {NoStop}%
\bibitem [{\citenamefont {Deffner}\ and\ \citenamefont
  {Saxena}(2015{\natexlab{a}})}]{Deffner2015PRL}%
  \BibitemOpen
  \bibfield  {author} {\bibinfo {author} {\bibfnamefont {Sebastian}\
  \bibnamefont {Deffner}}\ and\ \bibinfo {author} {\bibfnamefont {Avadh}\
  \bibnamefont {Saxena}},\ }\bibfield  {title} {\enquote {\bibinfo {title}
  {Jarzynski equality in $\mathcal{P}\mathcal{T}$-symmetric quantum
  mechanics},}\ }\href {\doibase 10.1103/PhysRevLett.114.150601} {\bibfield
  {journal} {\bibinfo  {journal} {Phys. Rev. Lett.}\ }\textbf {\bibinfo
  {volume} {114}},\ \bibinfo {pages} {150601} (\bibinfo {year}
  {2015}{\natexlab{a}})}\BibitemShut {NoStop}%
\bibitem [{\citenamefont {Deffner}\ and\ \citenamefont
  {Saxena}(2015{\natexlab{b}})}]{Deffner2015PRE}%
  \BibitemOpen
  \bibfield  {author} {\bibinfo {author} {\bibfnamefont {Sebastian}\
  \bibnamefont {Deffner}}\ and\ \bibinfo {author} {\bibfnamefont {Avadh}\
  \bibnamefont {Saxena}},\ }\bibfield  {title} {\enquote {\bibinfo {title}
  {Quantum work statistics of charged dirac particles in time-dependent
  fields},}\ }\href {\doibase 10.1103/PhysRevE.92.032137} {\bibfield  {journal}
  {\bibinfo  {journal} {Phys. Rev. E}\ }\textbf {\bibinfo {volume} {92}},\
  \bibinfo {pages} {032137} (\bibinfo {year} {2015}{\natexlab{b}})}\BibitemShut
  {NoStop}%
\bibitem [{\citenamefont {Talkner}\ and\ \citenamefont
  {H{\"{a}}nggi}(2016)}]{Talkner16}%
  \BibitemOpen
  \bibfield  {author} {\bibinfo {author} {\bibfnamefont {Peter}\ \bibnamefont
  {Talkner}}\ and\ \bibinfo {author} {\bibfnamefont {Peter}\ \bibnamefont
  {H{\"{a}}nggi}},\ }\bibfield  {title} {\enquote {\bibinfo {title} {Aspects of
  quantum work},}\ }\href
  {https://journals.aps.org/pre/abstract/10.1103/PhysRevE.93.022131} {\bibfield
   {journal} {\bibinfo  {journal} {Phys. Rev. E}\ }\textbf {\bibinfo {volume}
  {93}},\ \bibinfo {pages} {022131} (\bibinfo {year} {2016})}\BibitemShut
  {NoStop}%
\bibitem [{\citenamefont {Gardas}\ \emph {et~al.}(2016)\citenamefont {Gardas},
  \citenamefont {Deffner},\ and\ \citenamefont {Saxena}}]{Gardas2016}%
  \BibitemOpen
  \bibfield  {author} {\bibinfo {author} {\bibfnamefont {Bart{\l}omiej}\
  \bibnamefont {Gardas}}, \bibinfo {author} {\bibfnamefont {Sebastian}\
  \bibnamefont {Deffner}}, \ and\ \bibinfo {author} {\bibfnamefont {Avadh}\
  \bibnamefont {Saxena}},\ }\bibfield  {title} {\enquote {\bibinfo {title}
  {Non-hermitian quantum thermodynamics},}\ }\href {\doibase 10.1038/srep23408}
  {\bibfield  {journal} {\bibinfo  {journal} {Scientific Reports}\ }\textbf
  {\bibinfo {volume} {6}},\ \bibinfo {pages} {23408} (\bibinfo {year}
  {2016})}\BibitemShut {NoStop}%
\bibitem [{\citenamefont {Bartolotta}\ and\ \citenamefont
  {Deffner}(2018)}]{Bartolotta2018}%
  \BibitemOpen
  \bibfield  {author} {\bibinfo {author} {\bibfnamefont {Anthony}\ \bibnamefont
  {Bartolotta}}\ and\ \bibinfo {author} {\bibfnamefont {Sebastian}\
  \bibnamefont {Deffner}},\ }\bibfield  {title} {\enquote {\bibinfo {title}
  {Jarzynski equality for driven quantum field theories},}\ }\href {\doibase
  10.1103/PhysRevX.8.011033} {\bibfield  {journal} {\bibinfo  {journal} {Phys.
  Rev. X}\ }\textbf {\bibinfo {volume} {8}},\ \bibinfo {pages} {011033}
  (\bibinfo {year} {2018})}\BibitemShut {NoStop}%
\bibitem [{\citenamefont {Gardas}\ and\ \citenamefont
  {Deffner}(2018)}]{Gardas2018}%
  \BibitemOpen
  \bibfield  {author} {\bibinfo {author} {\bibfnamefont {Bart{\l}omiej}\
  \bibnamefont {Gardas}}\ and\ \bibinfo {author} {\bibfnamefont {Sebastian}\
  \bibnamefont {Deffner}},\ }\bibfield  {title} {\enquote {\bibinfo {title}
  {Quantum fluctuation theorem for error diagnostics in quantum annealers},}\
  }\href {\doibase 10.1038/s41598-018-35264-z} {\bibfield  {journal} {\bibinfo
  {journal} {Scientific Reports}\ }\textbf {\bibinfo {volume} {8}},\ \bibinfo
  {pages} {17191} (\bibinfo {year} {2018})}\BibitemShut {NoStop}%
\bibitem [{\citenamefont {Touil}\ and\ \citenamefont
  {Deffner}(2021)}]{Touil2021PRQ}%
  \BibitemOpen
  \bibfield  {author} {\bibinfo {author} {\bibfnamefont {Akram}\ \bibnamefont
  {Touil}}\ and\ \bibinfo {author} {\bibfnamefont {Sebastian}\ \bibnamefont
  {Deffner}},\ }\bibfield  {title} {\enquote {\bibinfo {title} {Information
  scrambling versus decoherence---two competing sinks for entropy},}\ }\href
  {\doibase 10.1103/PRXQuantum.2.010306} {\bibfield  {journal} {\bibinfo
  {journal} {PRX Quantum}\ }\textbf {\bibinfo {volume} {2}},\ \bibinfo {pages}
  {010306} (\bibinfo {year} {2021})}\BibitemShut {NoStop}%
\bibitem [{\citenamefont {Nielsen}\ and\ \citenamefont
  {Chuang}(2011)}]{Nielsen11}%
  \BibitemOpen
  \bibfield  {author} {\bibinfo {author} {\bibfnamefont {M.~A.}\ \bibnamefont
  {Nielsen}}\ and\ \bibinfo {author} {\bibfnamefont {I.~L.}\ \bibnamefont
  {Chuang}},\ }\href
  {https://www.cambridge.org/core/books/quantum-computation-and-quantum-information/01E10196D0A682A6AEFFEA52D53BE9AE}
  {\emph {\bibinfo {title} {Quantum Computation and Quantum Information: 10th
  Anniversary Edition}}},\ \bibinfo {edition} {10th}\ ed.\ (\bibinfo
  {publisher} {Cambridge University Press},\ \bibinfo {address} {New York, NY,
  USA},\ \bibinfo {year} {2011})\BibitemShut {NoStop}%
\bibitem [{\citenamefont {{\L}obejko}(2021)}]{Lobejko21}%
  \BibitemOpen
  \bibfield  {author} {\bibinfo {author} {\bibfnamefont {Marcin}\ \bibnamefont
  {{\L}obejko}},\ }\bibfield  {title} {\enquote {\bibinfo {title} {The tight
  second law inequality for coherent quantum systems and finite-size heat
  baths},}\ }\href {https://www.nature.com/articles/s41467-021-21140-4}
  {\bibfield  {journal} {\bibinfo  {journal} {Nat. Commun.}\ }\textbf {\bibinfo
  {volume} {12}},\ \bibinfo {pages} {918} (\bibinfo {year} {2021})}\BibitemShut
  {NoStop}%
\bibitem [{\citenamefont {Baumgratz}\ \emph {et~al.}(2014)\citenamefont
  {Baumgratz}, \citenamefont {Cramer},\ and\ \citenamefont
  {Plenio}}]{Baumgratz14}%
  \BibitemOpen
  \bibfield  {author} {\bibinfo {author} {\bibfnamefont {T.}~\bibnamefont
  {Baumgratz}}, \bibinfo {author} {\bibfnamefont {M.}~\bibnamefont {Cramer}}, \
  and\ \bibinfo {author} {\bibfnamefont {M.B.}\ \bibnamefont {Plenio}},\
  }\bibfield  {title} {\enquote {\bibinfo {title} {Quantifying coherence},}\
  }\href {https://journals.aps.org/prl/abstract/10.1103/PhysRevLett.113.140401}
  {\bibfield  {journal} {\bibinfo  {journal} {Phys. Rev. Lett.}\ }\textbf
  {\bibinfo {volume} {113}},\ \bibinfo {pages} {140401} (\bibinfo {year}
  {2014})}\BibitemShut {NoStop}%
\bibitem [{\citenamefont {Ashtekar}\ and\ \citenamefont
  {Schilling}(1999)}]{Ashtekar99}%
  \BibitemOpen
  \bibfield  {author} {\bibinfo {author} {\bibfnamefont {Abhay}\ \bibnamefont
  {Ashtekar}}\ and\ \bibinfo {author} {\bibfnamefont {Troy~A.}\ \bibnamefont
  {Schilling}},\ }\href
  {https://link.springer.com/chapter/10.1007/978-1-4612-1422-9_3} {\emph
  {\bibinfo {title} {``Geometrical Formulation of Quantum Mechanics" in ``On
  Einstein’s Path" (pp. 23-65)}}}\ (\bibinfo  {publisher} {Springer},\
  \bibinfo {address} {New York, NY, USA},\ \bibinfo {year} {1999})\BibitemShut
  {NoStop}%
\bibitem [{\citenamefont {Bengtsson}\ and\ \citenamefont
  {Zyczkowski}(2017)}]{Zyczkowski17}%
  \BibitemOpen
  \bibfield  {author} {\bibinfo {author} {\bibfnamefont {I.}~\bibnamefont
  {Bengtsson}}\ and\ \bibinfo {author} {\bibfnamefont {K.}~\bibnamefont
  {Zyczkowski}},\ }\href
  {https://www.cambridge.org/core/books/geometry-of-quantum-states/4BA9DCEED5BB16B222A917EAAAD17028}
  {\emph {\bibinfo {title} {Geometry of Quantum States}}}\ (\bibinfo
  {publisher} {Cambridge University Press},\ \bibinfo {year}
  {2017})\BibitemShut {NoStop}%
\bibitem [{\citenamefont {Cari{\~{n}}ena}\ \emph {et~al.}(2007)\citenamefont
  {Cari{\~{n}}ena}, \citenamefont {Clemente-Gallardo},\ and\ \citenamefont
  {Marmo}}]{Marmo07}%
  \BibitemOpen
  \bibfield  {author} {\bibinfo {author} {\bibfnamefont {J.~F.}\ \bibnamefont
  {Cari{\~{n}}ena}}, \bibinfo {author} {\bibfnamefont {J.}~\bibnamefont
  {Clemente-Gallardo}}, \ and\ \bibinfo {author} {\bibfnamefont
  {G.}~\bibnamefont {Marmo}},\ }\bibfield  {title} {\enquote {\bibinfo {title}
  {Geometrization of quantum mechanics},}\ }\href {\doibase
  https://doi.org/10.1007/s11232-007-0075-3} {\bibfield  {journal} {\bibinfo
  {journal} {Theor. Math. Phys.}\ }\textbf {\bibinfo {volume} {152}},\ \bibinfo
  {pages} {894} (\bibinfo {year} {2007})}\BibitemShut {NoStop}%
\bibitem [{\citenamefont {Anza}\ and\ \citenamefont {Crutchfield}()}]{Anza21}%
  \BibitemOpen
  \bibfield  {author} {\bibinfo {author} {\bibfnamefont {Fabio}\ \bibnamefont
  {Anza}}\ and\ \bibinfo {author} {\bibfnamefont {James}\ \bibnamefont
  {Crutchfield}},\ }\href@noop {} {\bibinfo  {journal} {in preparation}\
  }\BibitemShut {NoStop}%
\bibitem [{\citenamefont {Deffner}\ and\ \citenamefont
  {Lutz}(2010)}]{Deffner2010PRL}%
  \BibitemOpen
\bibfield  {journal} {  }\bibfield  {author} {\bibinfo {author} {\bibfnamefont
  {Sebastian}\ \bibnamefont {Deffner}}\ and\ \bibinfo {author} {\bibfnamefont
  {Eric}\ \bibnamefont {Lutz}},\ }\bibfield  {title} {\enquote {\bibinfo
  {title} {Generalized clausius inequality for nonequilibrium quantum
  processes},}\ }\href {\doibase 10.1103/PhysRevLett.105.170402} {\bibfield
  {journal} {\bibinfo  {journal} {Phys. Rev. Lett.}\ }\textbf {\bibinfo
  {volume} {105}},\ \bibinfo {pages} {170402} (\bibinfo {year}
  {2010})}\BibitemShut {NoStop}%
\bibitem [{\citenamefont {Jaynes}(1957)}]{Jaynes57}%
  \BibitemOpen
  \bibfield  {author} {\bibinfo {author} {\bibfnamefont {E.~T.}\ \bibnamefont
  {Jaynes}},\ }\bibfield  {title} {\enquote {\bibinfo {title} {Information
  theory and statistical mechanics},}\ }\href {\doibase
  https://doi.org/10.1103/PhysRev.106.620} {\bibfield  {journal} {\bibinfo
  {journal} {Phys. Rev.}\ }\textbf {\bibinfo {volume} {106}},\ \bibinfo {pages}
  {620} (\bibinfo {year} {1957})}\BibitemShut {NoStop}%
\end{thebibliography}%

\end{document}